\documentclass[tog,preprint]{acmodified}
\pdfoutput=1

\usepackage{amsmath}
\usepackage{amssymb}

\usepackage{pgfplots}

\usepgfplotslibrary{units}
\usepackage{xparse}
\usepackage{xspace}

\newcommand{\symb}[1]{{\ensuremath{#1}}\xspace}

\DeclareDocumentCommand \argmax { s D[]{} m } {
	\IfBooleanTF{#1}
		  {\symb{\operatorname*{arg\,max}_{#2} #3}}
		  {\symb{\operatorname{arg\,max}_{#2} #3}}
}

\DeclareDocumentCommand \argmin { s D[]{} m } {
	\IfBooleanTF{#1}
		  {\symb{\operatorname*{arg\,min}_{#2} #3}}
		  {\symb{\operatorname{arg\,min}_{#2} #3}}
}

\DeclareDocumentCommand \newsymbol {m m D[]{} D<>{} d()} {
	\DeclareDocumentCommand #1 { t^ s d[] d<> d() }{%
		\symb{ %
		\IfBooleanTF{##2}%
		{\IfBooleanTF{##1}{\hat{#2}}{#2}%
			_{\IfNoValueTF{##3}{}{##3}}%
			^{\IfNoValueTF{##4}{}{##4}}%
			\IfNoValueTF{##5}{}{(##5)}
		}%
		{\IfBooleanTF{##1}{\hat{#2}}{#2}%
			_{\IfNoValueTF{##3}{#3}{##3}}%
			^{\IfNoValueTF{##4}{#4}{##4}}%
			\IfNoValueTF{##5}%
				{\IfNoValueTF{#5}{}{(#5)}}%
				{(##5)}%
		}}%
	}%
}

\newsymbol{\p}{p}()

\DeclareDocumentCommand \set { m d[] }{%
	\symb{\{#1\}_{{\IfNoValueTF{#2}{}{#2}}}}%
}

\newcommand{\eqn}[1]{\eqref{eq:#1}}
\newcommand{\eql}[1]{\label{eq:#1}}

\newcommand{\figl}[1]{\label{fig:#1}}

\newcommand{\secref}[1]{\S\ref{sec:#1}}
\newcommand{\secl}[1]{\label{sec:#1}}

\newcommand{\ie}{\textit{i.e.}\xspace}

\newcommand{\eg}{\textit{e.g.}\xspace}

\newcommand{\done}[1]{}
\newcommand{\elsewhere}[1]{}

\newcommand{\biglbrace}[1]{\left\lbrace\begin{minipage}[t]{0cm}\vspace{#1}\end{m
inipage}\right.}

\newsymbol{\domain}{\Omega}
\newsymbol{\cell}{\Omega}[\sample]
\newcommand{\sample}{\symb{s}}

\newcommand{\oidx}{\symb{o}}
\newcommand{\oidxsurf}{\symb{o}}
\newcommand{\oidxvol}{\symb{o}}
\newcommand{\vidx}{\symb{q}}
\newcommand{\vidxsurf}{\symb{v}}
\newcommand{\vidxvol}{\symb{s}}
\newsymbol{\vertnorm}{\vec{\mathbf{n}}}[\vidxsurf]<\tidx>
\newsymbol{\anglemaxnormal}{\theta}[\mathrm{max}]

\newcommand{\tidx}{\symb{t}}

\newsymbol{\ksel}{\mathrm{k}}[\oidx]<\tidx>
\newsymbol{\kselset}{\mathrm{K}}
\newsymbol{\kselalt}{\tilde{k}}
\newsymbol{\vselsurf}{\vidxsurf}
\newsymbol{\vselvol}{\vidxvol}
\newsymbol{\vselalt}{\tilde{\vidx}}
\newsymbol{\patchpose}{\mathbf{T}}[k]<\tidx>
\newsymbol{\patchrotation}{\mathbf{R}}[k]<\tidx>
\newsymbol{\patchposem}{\mathbf{\bar{T}}}[k]
\newsymbol{\distance}{\mathcal{D}}
\newsymbol{\M}{\mathbf{M}}[k,l] % inter-patch transform
\newsymbol{\patch}{P}[k]
\newsymbol{\patchsurf}{SP}[k]
\newsymbol{\patchneighborhood}{N}[k]
\newsymbol{\patchedgeset}{\mathcal{E}}
\newsymbol{\patchorientededgeset}{\overrightarrow{\mathcal{E}}}
\newsymbol{\patchpairset}{\mathcal{N}}
\newsymbol{\x}{\mathbf{x}}[\vidx]<t>
\newsymbol{\xref}{\mathbf{x}}[\vidx]<0>
\newsymbol{\xm}{\mathbf{\bar{x}}}[\vidx]
\newsymbol{\w}{\alpha}[k](v)
\newsymbol{\obs}{\mathbf{y}}[\oidx]<\tidx>
\newsymbol{\Obs}{\mathbf{Y}}
\newsymbol{\obssurf}{\mathbf{y}}[\oidxsurf]<\tidx>
\newsymbol{\obssurfnorm}{\vec{\mathbf{n}}}[\oidxsurf]<\tidx>
\newsymbol{\Obssurf}{\mathbf{Y}}
\newsymbol{\obsvol}{\mathbf{y}}[\oidxvol]<\tidx>
\newsymbol{\obsvoldist}{\mathbf{d}}[\oidxvol]<\tidx>
\newsymbol{\Obsvol}{\mathbf{Y}}
\newsymbol{\distancetolerance}{\epsilon}
\newsymbol{\E}{E}

\newsymbol{\noise}{{\sigma}}<\tidx>
\newsymbol{\noiseestim}{{{\hat{\sigma}}_0}}
\newsymbol{\rigidc}{\mathrm{c}}[kl]
\newsymbol{\rigidcset}{\mathrm{C}}
\newsymbol{\springlength}{d}
\newsymbol{\timewindow}{\mathcal{T}}
\newsymbol{\oidxset}{\mathcal{O}}[\tidx]
\newsymbol{\oidxsetsurf}{\mathcal{SO}}[\tidx]
\newsymbol{\oidxsetvol}{\mathcal{VO}}[\tidx]
\newsymbol{\vertset}{\mathcal{S}}
\newsymbol{\vertsetvol}{\mathcal{C}}
\newsymbol{\vertsetsurf}{\mathcal{V}}
\newsymbol{\param}{\Theta}
\newsymbol{\Z}{Z}

\newcommand{\runner}{\textsc{Runner}}
\newcommand{\cagebirddance}{\textsc{BirdCageDance}}
\newcommand{\slackline}{\textsc{Slackline}}

\newsymbol{\heat}{\mathbf{F}}[t]
\newsymbol{\heatkernel}{\mathbf{H}}[t]
\newsymbol{\eigenvect}{\mathbf{V}}[k]
\newsymbol{\eigenval}{\mathbf{\lambda}}[k]
\newsymbol{\laplacian}{\mathbf{L}}

\title{Shape Animation with Combined Captured and Simulated Dynamics}

\author{Benjamin Allain, Li Wang, Jean-S\'ebastien Franco, Franck Hetroy-Wheeler, and Edmond Boyer\\
 LJK-INRIA Grenoble Rh\^one-Alpes, France}

\pdfauthor{Benjamin Allain, Li Wang, Jean-Sébastien Franco, Franck Hetroy and Edmond Boyer}

\begin{document}

\teaser{
   \includegraphics[width=0.95\linewidth]{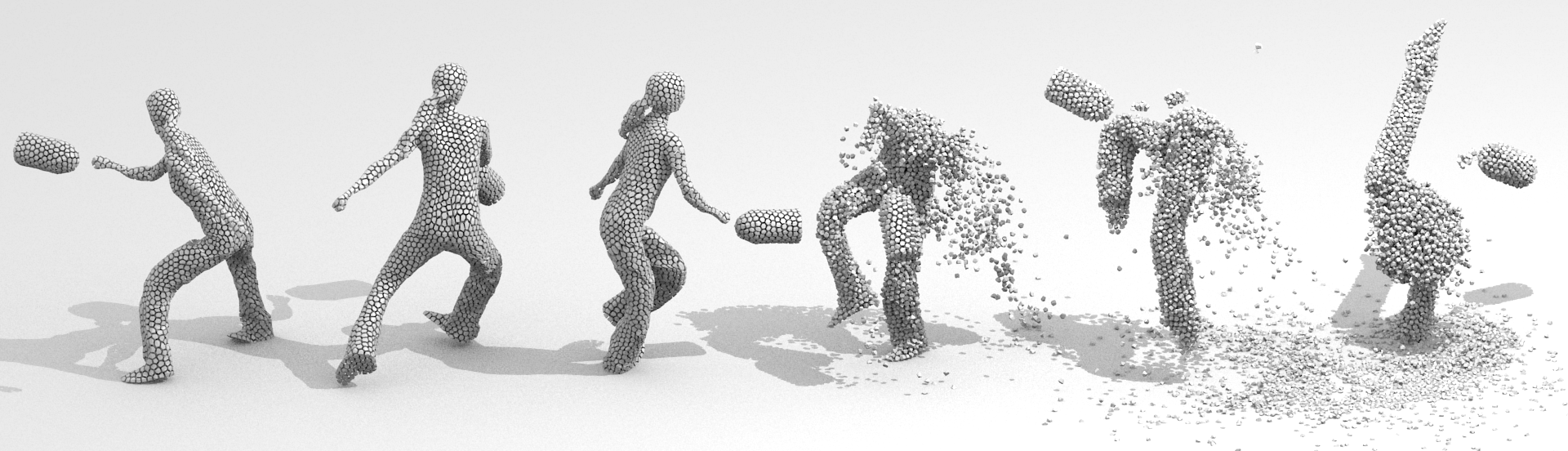}
 \caption{An animation that combines video-based shape motion (left) and physical simulation (right). Our method allows to apply mechanical effects on captured dynamic shapes and generates therefore  plausible animations with real dynamics.}  
  \figl{teaser}
}

\maketitle

\begin{abstract}
We present a novel volumetric animation generation framework to create new types of animations from raw 3D surface or point cloud sequence of captured real performances.  The framework considers as input time incoherent 3D observations of a moving shape, and is thus particularly suitable for the output of performance capture platforms. In our system, a suitable virtual representation of the actor is built from real captures that allows seamless combination and simulation with virtual external forces and objects, in which the original captured actor can be reshaped, disassembled or reassembled from user-specified virtual physics. Instead of using the dominant surface-based geometric representation of the capture, which is less suitable for volumetric effects, our pipeline exploits Centroidal Voronoi tessellation decompositions as unified volumetric representation of the real captured actor, which we show can be used seamlessly as a building block for all processing stages, from capture and tracking to virtual physic simulation. The representation makes no human specific assumption and can be used to capture and re-simulate the actor with props or other moving scenery elements. We demonstrate the potential of this pipeline for virtual reanimation of a real captured event with various unprecedented volumetric visual effects, such as volumetric distortion, erosion, morphing, gravity pull, or collisions.

\end{abstract}

%-------------------------------------------------------------------------
\section{Introduction}

Creation of animated content has become of major interest for many applications, notably in the entertainment industry, where the ability to produce animated virtual characters is central to video games and special effects. Plausibility of the animations is a significant concern for such productions, as they are critical to the immersion and perception of the audience. Because of the inherent difficulty and necessary time required to produce such plausible animations from scratch, motion capture technologies are now extensively used to obtain kinematic motion data as a basis to produce the animations, and are now standard in the industry.

However, motion capture is usually only the first stage in a complicated process, before the final animation can be obtained. The task requires large amounts of manual labor to rig the kinematic data to a surface model, correct and customize the animation, and produce the specifics of the desired effect. This is why, in recent times, video-based 3D performance capture technologies are gaining more and more attention, as they can be used to directly produce 3D surface animations with more automation, and to circumvent many intermediate stages in this process. They also make it possible to automatically acquire complex scenes, shapes and interactions between characters that may not be possible with the standard sparse-marker capture technologies. Still, the problem of customizing the surface animations produced by such technologies to yield a modified animation or a particular effect has currently no general and widespread solution, as it is a lower level representation to begin with. %FRANCK: je comprends pas cette phrase : qu'est-ce qui est une représentation bas niveau ?

In this work, we propose a novel system towards this goal, which produces animations from a stream of 3D observations acquired with a video-based capture system. The system provides a framework to push the automation of animation generation to a new level, dealing with the capture, shape tracking, and animation generation from end-to-end with a unified representation and solution. In particular, we entirely circumvent the need for kinematic rigging and present results in this report obtained without any manual surface correction.

Although the framework opens many effect possibilities, for the purpose of the demonstration here, we focus our effort on combining the real raw surface data captured with physics and procedural animation, in particular using a physics-based engine. To this aim, we propose to use regular Voronoi tessellations to decompose acquired shapes into volumetric cells, as a dense volume representation upon which physical constraints are easily combined with the captured motion constraints. Hence, shape motions can  be perturbed with various effects in the animation, through forces or procedural decisions applied on volumetric cells. Motion constraints are obtained from captured multi-view sequences of live actions. We do not consider skeletal or surfacic motion models for that purpose but directly track volumetric cells instead. This ensures high flexibility in both the class of shapes and the class of physical constraints that can be accounted for. We have evaluated our method with various actor performances and effects. We provide both quantitative results for the shape tessellation approach and qualitative results for the generated 3D content. They demonstrate that convincing and, to the best of our knowledge, unprecedented animations can be obtained using video-based captured shape models. 
 
 In summary, this work considers video-based animation and takes the field a step further by allowing for physics-based or procedural animation effects. The core innovation that permits the combination of real and simulated dynamics lies in the volumetric shape representation we propose. The associated tessellated  volumetric cells can be both tracked and physically perturbed hence enabling new computer animations.

%FRANCK: I would say: the motivation is to allow easier/more realistic character animation (more than just characters? In this case reviewers will expect fluids). State %of the art only allows:
%\begin{itemize}
%\item either animation/deformation of a static shape;
%\item or tracking of a dynamic shape, but without an easy way to modify the recorded motion (because often only sparse motion information).
%\end{itemize}
%We propose to combine dense real shape in motion data and physical constraints to be able to deform a given recorded dynamic shape in a user-friendly manner. Why: interest in the media industry, but not only (cite others). To reach this goal we propose to take advantage of multi-views platforms that now allow to capture complex dense dynamic scenes. Specifically, we propose a volumetric shape representation that allows both dense surface tracking and physical perturbation.

%-------------------------------------------------------------------------
\section{Related work}

%mention that only video-based are cited below ?

%Mocap combination with physical simulation
This work deals with the combination of simulated and captured shape motion data. As mentioned earlier, this has already been explored with marker based mocap data as kinematic constraints. Following the work of \cite{popovic99},  a number of researchers have investigated such combination. They propose methods where mocap data can be used either as reference motion~\cite{popovic99,zordan02,sulejmanpasic05}, or to constrain the  physics-based optimization associated to the simulation with human-like motion styles~\cite{liu05,safonova04,ye08,wei11}. Although sharing conceptual similarities with these methods, our work differs substantially. Since  video-based animations already provide natural animations, our primary objective is not to constrain a physical model with captured kinematic constraints but rather to enhance captured animations with user-specified animation constraints based on physics or procedural effects. Consequently, our simulations are not based on biomechanical models but on  dynamic simulations of mechanical effects. Nevertheless, our research draws inspiration from these works. 

%creating new animation by edition, motion transfer 
With the aim to create new animations using recorded video-based animations,  some works consider the concatenation of elementary segments of animations, \eg~\cite{casas13}, the local deformation of a given animation, \eg~\cite{cashman12}, or the transfer of a deformation between captured surfaces, \eg~\cite{sumner04}. While we also aim at generating new animations, we tackle a different issue in this research with the perturbation of recorded  animations according to simulated effects. 

%History
Our method builds on results obtained in video-based animations with multi-camera setups, to obtain the input data of our system. Classically, multi-view silhouettes can be used to build free viewpoint videos using visual hulls~\cite{matusik00,gross03} or to fit a synthetic body model~\cite{carranza03}. Visual quality of reconstructed shape models can be improved by considering photometric information~\cite{starck07,tung09} and also by using laser scanned models as templates that are deformed and tracked over temporal sequences~\cite{aguiar07,vlasic08,aguiar08}. Interestingly, these shape tracking strategies provide temporally coherent 3D models that carry therefore motion information. In addition to geometric and photometric information, considering shading cues allows to recover finer scale surface details as in~\cite{vlasic09,wu12} .  

%shape and motion recovery-> our objective
More recent approaches have proposed to recover both shapes and motions. They follow various directions depending on the prior information assumed for shapes and their deformations. For instance in the case of human motion, a body of work assumes articulated motions that can be represented by the poses of  skeleton based models, \eg~\cite{vlasic08,gall09,straka12}. We base our system on a different class of techniques aiming at more general scenarios, with less constrained motion models   simply based on locally rigid assumptions in the shape volume~\cite{allain15}. This has the advantage that a larger class of shapes and deformations can be considered, in particular motions of humans with loose clothes or props. The technique also has the significant advantage that it allows to track dense volumetric cell decompositions of objects, thereby allowing for consistent captured motion information to be associated and propagated with each cell in the volume, a key property to build our animation generation framework on.

In the following, we provide a system overview, followed by a detailed explanation of how we tessellate 3D input observations into regular polyhedral cells, to be subsequently tracked and used as primary animation entity (\secref{sec:cvt}). In order to recover kinematic constraints from real actions, our system then tracks polyhedral cells using surface observations and a locally rigid deformation model (\secref{sec:tracking}). Finally, a physics or procedural simulation integrates the animation constraints over the shape (\secref{sec:anim}). To our knowledge, this is the first attempt to propose such an end-to-end system and framework to generate animations from real captured dynamic shapes.

%Physical based modeling: not the contribution of the paper. Easier over volumetric cells and with existing tools.

%Combination of both : new to the best of our knowledge with dense shape models. Do we cite here works doing this with skeletons ?

%FRANCK: some papers that we may cite:
%\begin{itemize}
%\item Casas TVCG 2013: deformation of a dynamic shape, but limited (no topology change for instance)
%\item Cashman/Hormann EG 2012: deformation of a dynamic shape, limited (temporally consistent mesh sequence as input, deformation with same topology and connectivity)
%\item Kircher/Garland ToG 2008: limited, time consuming
%\item Sumner/Schmid/Pauly SIGGRAPH 2007: space deformation method, limited (no topology change)
%\end{itemize}

%-------------------------------------------------------------------------
\section{System Overview}

\begin{figure*}[htpb]
%\centering
\hspace*{-3mm}
\includegraphics[width=\linewidth]{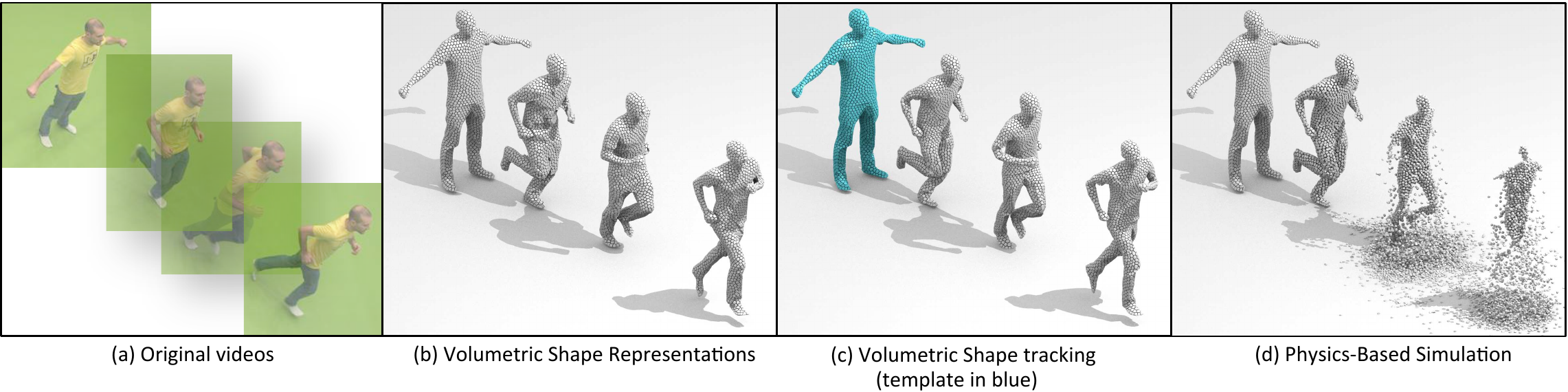}
%\begin{tabular} {cccc}
%	\begin{tabular}{cc}
%	\includegraphics[width=.07\linewidth]{imagesEG/overview/img-000000} &
%	\includegraphics[width=.07\linewidth]{imagesEG/overview/img-000164} \\
%	\includegraphics[width=.07\linewidth]{imagesEG/overview/img-000191} &
%	\includegraphics[width=.07\linewidth]{imagesEG/overview/img-000215} 
%	\end{tabular} &
%	\includegraphics[width=.25\linewidth]{imagesEG/overview/overviewA} &
% 	\includegraphics[width=.25\linewidth]{imagesEG/overview/overviewB} &
% 	\includegraphics[width=.25\linewidth]{imagesEG/overview/overviewC}
%\end{tabular}
\caption{From video-based shape capture to physic simulation. The approach uses multiple videos and  Voronoi tessellations to capture the volumetric kinematic of a shape motion which can then be reanimated with additional mechanical effects, for instance volumetric erosion with gravity in the figure. }
\figl{outline}
\end{figure*}

We generate a physically plausible animation given a sequence of 3D shape observations as well as user specifications for the desired effect to be applied on the animation. 3D observations are  transformed into temporally consistent volumetric models using centroidal Voronoi tessellations and shape tracking. Kinematic and physical constraints are then combined using rigid body physics simulation. The approach involves the following main steps depicted in Figure~\ref{fig:outline}.

\textbf{Video based acquisition} 
Input to our system are 3D observations of  a dynamic scene. Traditional and probably  most common dynamic scenes in graphics are composed of human movements; however our system can consider a larger class of shapes since only local rigidity is assumed to get temporally consistent shape models. 3D observations are assumed to be obtained using a multi-camera system and can be in any explicit, \eg meshes or implicit, \eg from point clouds through Poisson function, forms. Our own apparatus is composed of $68$ calibrated and synchronized cameras with a resolution up to $2048\times2048$ pixels. The acquisition space is about $8m\times4m$ and the camera frame rate can go up to $50$ fps at full resolution. The outputs of this step are point clouds  with around $100k$ points.

\textbf{Volumetric Representation.}
Input 3D observations are tessellated  into polyhedral cells. This volumetric representation is motivated by two aspects of our animation goal: first, the representation is well suited to physical simulation; second,   volumetric deformation models are more flexible than skeleton based models, hence enabling  non rigid shape deformations.  Still,  they allow for locally rigid volumetric constraints, a missing feature with surface deformation models when representing shapes that are volumes.  We adopt centroidal Voronoi tessellations that produce regular and uniform polyhedral cells. Several methods have been suggested to clip a Voronoi tessellation to a given surface, \eg~\cite{yan13,levy14}. However, to the best of our knowledge, none is able to handle point clouds without explicit neighboring information. In section~\secref{sec:cvt} we present a novel clipping method to compute CVT given an indicator function that identifies the two regions inside and outside the shape considered. Such an indicator function can be defined by an implicit function over a point cloud, \eg a Poisson function, or by an explicit form, \eg a mesh.

\textbf{Tracking.}
%note that even with temporally consistent meshes tracking over cells is still required
In this step, incoherent  volumetric shape models of a temporal sequence are transformed into coherent representations where a single shape model is evolving over time. This provides kinematic information at the cell level that will further be used in the simulation. We use a tracking method~\cite{allain15} that finds the poses of a given template shape at each frame. The template shape is taken as one of the  volumetric  models at a  frame.  The approach uses a volumetric deformation model, instead of surface or skeleton based model, to track shapes. It optimizes the pose of the template shape cells so as to minimize a distance cost to an input shape model while enforcing rigidity constraints on the local cell configurations. 
%It should be noticed here that using a skeletal or surface  based approach first to track shapes would not necessarily help since  inferring cell dynamics from a tracked skeleton  or a tracked surface amounts to track cells anyway. 

\textbf{Simulation.}
The tracked cell representation is both suitable for tracking and convenient for solid based physics. We embed the tracked volumetric model in a physical simulation, by considering each cell to be a rigid solid object in mechanical interaction with other cells and scene objects. We ensure cohesion of cells by attaching a kinematic recall force in the simulation, and offer various controls as to how the scene may deform, collide, or rupture during contacts and collisions. This simple framework allows for a number of interesting effects demonstrated in \secref{results}.

\section{Volumetric Shape Modeling}
\secl{sec:cvt}

In order to  perturb captured moving shapes with simulated mechanical effects, we resort to volumetric  discretizations. They enable  combined kinematic and physical constraints to be be applied over cells using rigid body simulations. To this goal, we partition shapes into volumetric cells using Voronoi  tessellations. Ideally,  cells should be regular and uniform to ease the implementation of local constraints such as physical constraints for simulation or  local deformation constraints  when tracking shapes over time sequences. Volumetric voxel grids~\cite{lc87,ju02}, while efficient,  are biased towards the grid axes and can therefore produce tessellations with poor quality. Other solutions such as Delaunay tetrahedrizations, \eg~\cite{shewchuk98,jamin14}, can be considered  however they can present badly shaped cells such as slivers. Moreover, they can not always guarantee a correct topology for the output mesh since the boundary of a tetrahedral structure can always present non manifold parts. In this work, we consider  Centroidal Voronoi tessellations (CVTs) to model shapes and their evolutions. Resulting cells in CVTs are known to be uniform, compact, regular and isotropic~\cite{Du99} . In the following, we explain how to build CVT representations given the indicator function of a 3D shape. 

% \eg~\cite{furukawa10,microsoft15}. 

\subsection{Mathematical Background}
\label{ssec:cvt_math}
Given a finite set of $n$ points $X = \{x_i\}_{i=1}^n$, called \emph{sites}, in a $3$-dimensional Euclidean space $\mathbb{E}^3$, the \emph{Voronoi cell} or \emph{Voronoi region} $\Omega_i$ \cite{Oka00} of $x_i$ is defined as follows:
\[
\Omega_i = \{x\in \mathbb{E}^3 \ | \ \|x - x_i\| \leq \|x - x_j\|, \ \forall j \neq i\}.
\]
The partition of $\mathbb{E}^3$ into Voronoi cells is called a \emph{Voronoi tessellation}.

A \emph{clipped Voronoi tessellation} \cite{yan13} is the intersection between the Voronoi tessellation and a volume $\Omega$, bounded by the surface $S$. A \emph{clipped Voronoi cell} is thus defined as:
\[
\Omega_i = \{x\in \Omega \ | \ \|x - x_i\| \leq \|x - x_j\|, \ \forall j \neq i\}.
\]

A \emph{centroidal Voronoi tessellation} (CVT)~\cite{Du99} is a special type of clipped Voronoi tessellation where the site of each Voronoi cell is also its centre of mass. Let the clipped Voronoi cell $\Omega_i$ be endowed with a density function $\rho$ such that $\rho(x) > 0 ~\forall x \in \Omega_i$. The centre of mass ${x}_i$, also called the centroid, of $\Omega_i$ is then defined as follows:
\[
{x}_i = \frac{\int_{\Omega_i}\rho(x)x\,\mathrm{d}\sigma}{\int_{\Omega_i}\rho(x)\,\mathrm{d}\sigma},
\]
where $\mathrm{d}\sigma$ is the area differential.

CVTs are widely used to discretize 2D or 3D regions. In this respect, CVTs are optimal quantisers that minimise a distortion or quantization error defined as: 
\begin{equation}
E(X) = \sum_{i=1}^n F_i(X) = \sum_{i=1}^n \int_{\Omega_i}\rho(x)\|x-{x}_i\|^2\,\mathrm{d}\sigma.
\end{equation}
CVTs correspond to local minima of the above function $E$, also called the CVT energy function \cite{Du99}.
%
%Thanks to Eq.~(\ref{eq:energy}), the partial derivative of the CVT energy function with respect to each site is:
%\begin{equation}
%\label{eq:partial}
%\frac{\partial E}{\partial x_i} = 2 V_i (x_i - \hat{x}_i).
%\end{equation}
%where $V_i = \int_{\Omega_i}\rho(x)\sigma$.

\subsection{Algorithm}
\label{ssec:cvt_alg}

Our CVT computation algorithm shares the same pipeline as other CVT computation methods, except it takes as input a shape $\Omega$ whose boundary surface $S$ is not necessarily explicitly known. In our case, $S$ can be defined explicitly as a mesh, or implicitly over a point cloud with a function that can be either given or estimated, \eg a Poisson function.
From a prescribed number $n$ of sites, our algorithm consists of the following three main steps:
\begin{enumerate}
\item Initialization: find initial positions for the $n$ sites inside $S$.
\item Clipping: compute the Voronoi tessellation of the sites, then restrict it to $\Omega$ by computing its intersection with $S$.
\item Optimization: update the position of the sites by minimizing the CVT energy function.
\end{enumerate}
Steps $2$ and $3$ are iterated several times. The number of iterations is a user-defined parameter. 

%The initial positions of sites has a strong influence on the convergence speed and on the result quality, and there are several CVT initialisation methods such as random sampling method, Ward's method and Hammersley sampling method can be applied. Since the initialisation is not a part of contribution of this paper, we use random sampling method because it is fast and simple to implemented. The idea is to sample the initial site positions randomly inside the implicit surface $S$.

\paragraph*{1. Initialization} Any initialization can be applied in our framework. In our experiments, we have randomly positioned the sites inside $S$. These experiments show that such initialization is sufficient to generate better representations  than voxel based or Delaunay techniques (see Figure~\ref{fig:cvt2} and Table~\ref{tab:cvt}).

\paragraph*{2. Clipping} 
%Several methods have been suggested to clip a Voronoi tessellation to a given surface~\cite{yan13,levy14,wang15}. However, to the best of our knowledge, none is able to handle implicit surfaces. %TODO: vérifier
We have designed an efficient algorithm to compute the clipped Voronoi tessellation of a volumetric shape $V$ bounded by an implicit surface $S$. Given a 3D Voronoi tessellation $\displaystyle\bigcup_i \Omega_i$ with sites $\{x_i\}$ inside $V$, the algorithm proceeds as follows:

\begin{enumerate}
\item Identify the boundary Voronoi cells $\Omega_i$ which intersect the implicit surface $S$.
\item For each of these cells,
\begin{enumerate}
\item Compute its intersection with $S$. This intersection is represented by a set of points $P_i$ obtained by discretizing the facets of $\Omega_i$, and  its edges, and identifying the resulting discretized cells that intersect $S$.  
\item Construct the boundary clipped Voronoi cell $\Omega'_i$ by computing the convex hull of the union of all points in $P_i$ and the vertices of $\Omega_i$ inside $V$.
\end{enumerate}
\end{enumerate}

To identify the boundary cells, all infinite Voronoi cells are first converted to finite cells. This is done by replacing the infinite rays edging the cell by finite length segments, with a length greater than the diameter of a bounding sphere containing the shape. This creates new vertices for adjacent cells. A Voronoi cell is then detected as a boundary cell if at least one of its vertices is outside the shape.

\begin{figure}[htpb]
\centering
\begin{tabular} {cc}
  \includegraphics[width=.45\linewidth]{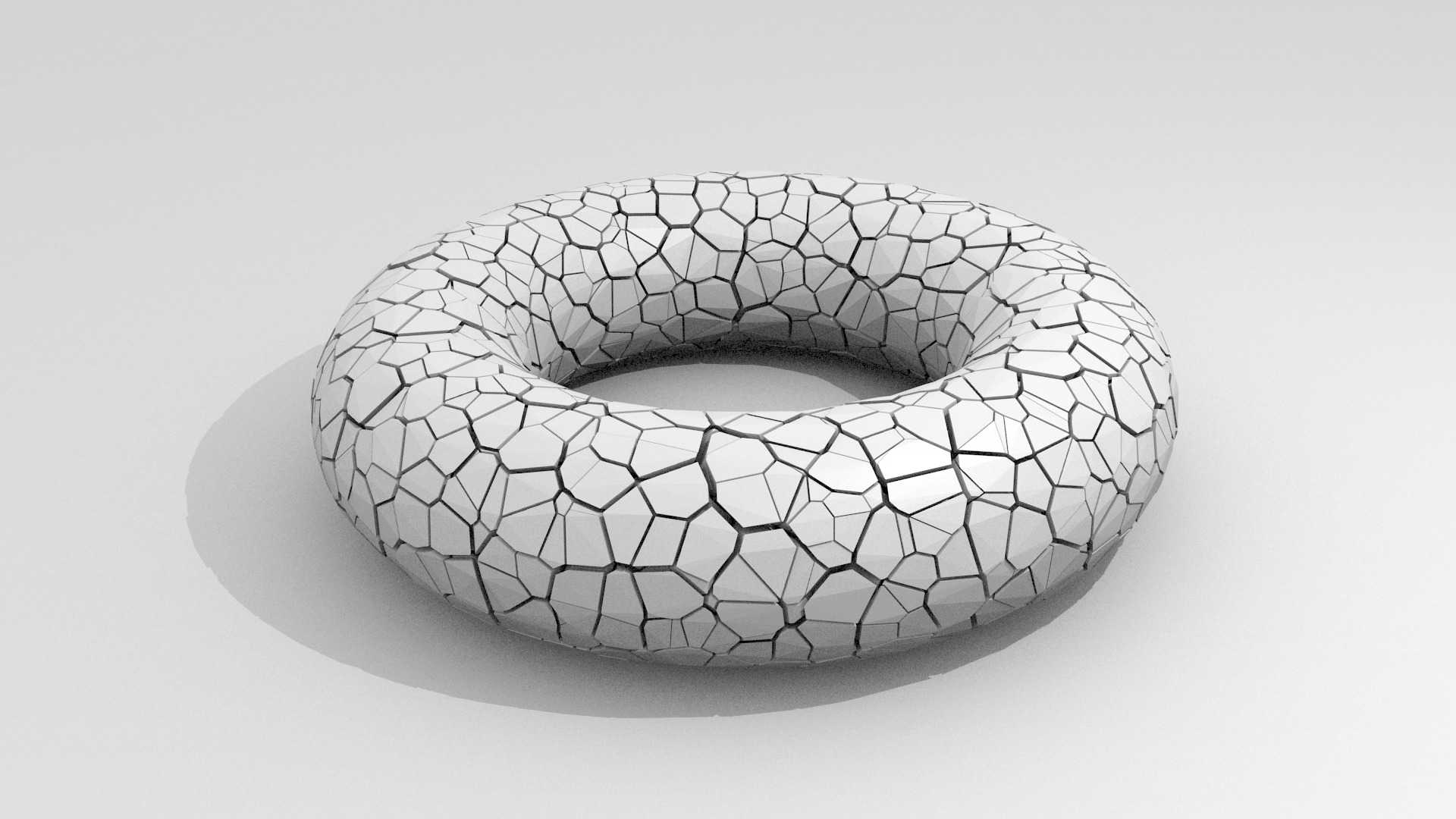} &
  \includegraphics[width=.45\linewidth]{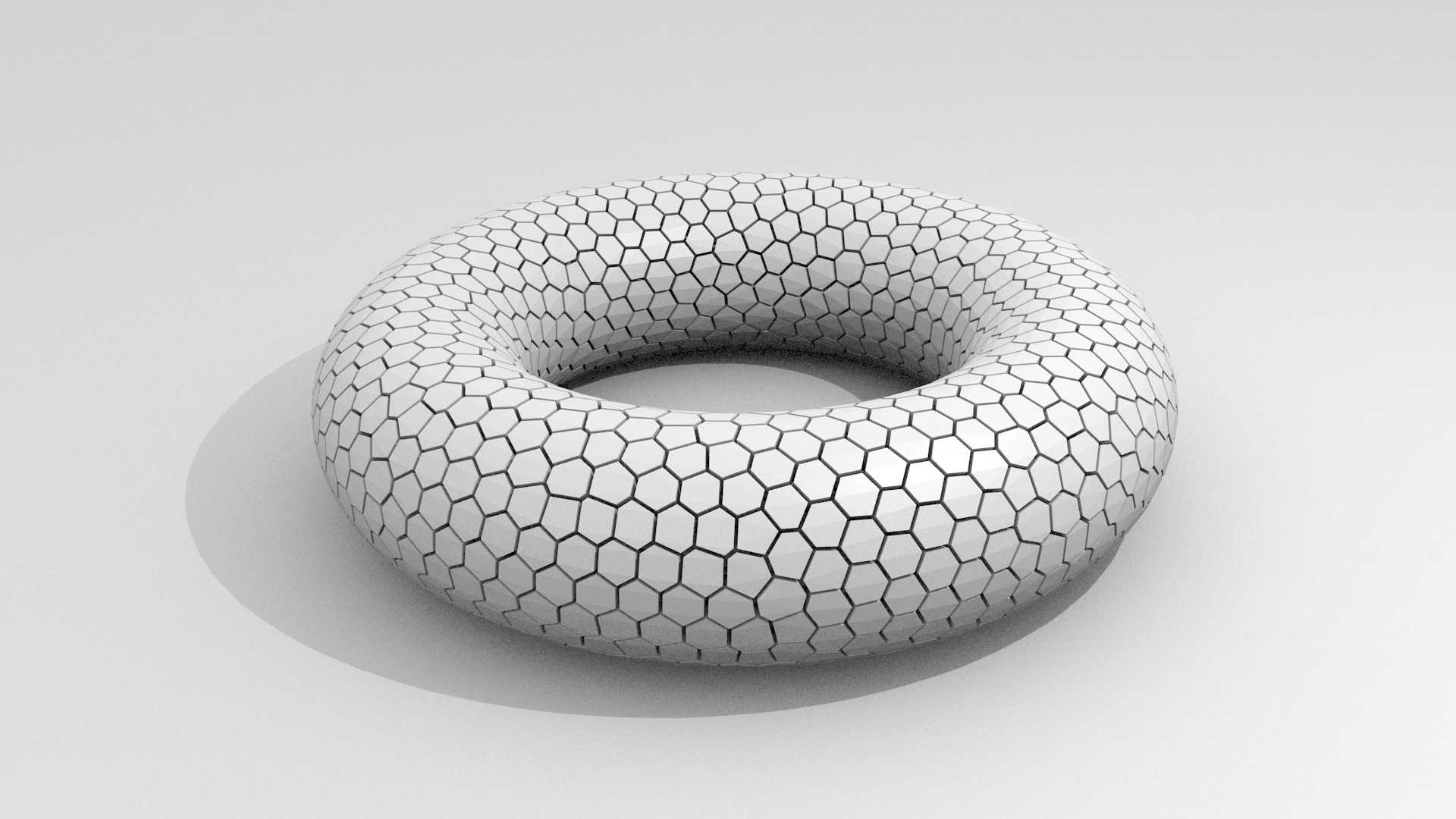} \\
  	(a) & (b) \\ \\
\end{tabular}
\caption{
	Clipped Voronoi tessellations without (a) and with optimized site locations (b). Cells in  displayed CVTs are slightly shrinked for visualization purposes.} \label{fig:cvt1}
\end{figure}

\paragraph*{3. Optimization}
Since the sites are not regularly distributed initially, the cells of the clipped Voronoi tessellation, as obtained with step 2, are not uniform nor regular, as shown in  Figure~\ref{fig:cvt1} (a). In order to improve cell shapes and to get a uniform and regular volumetric decomposition of $V$, the site locations are optimized. In the literature, the two main strategies for this optimization are the Lloyd's gradient descent method and the L-BFGS quasi-Newton method. In our approach, we choose the latter for its fast convergence~\cite{liu09}. As shown in Figure~\ref{fig:cvt1} (b), once convergence in the optimization is reached, all the clipped Voronoi cells present more uniform and regularly distributed shapes. 

\subsection{Evaluation}
\label{ssec:cvt_res}
\begin{figure*}[htpb]
%\centering
\hspace*{-4mm}
\begin{tabular} {cccc}
	\includegraphics[width=.195\linewidth]{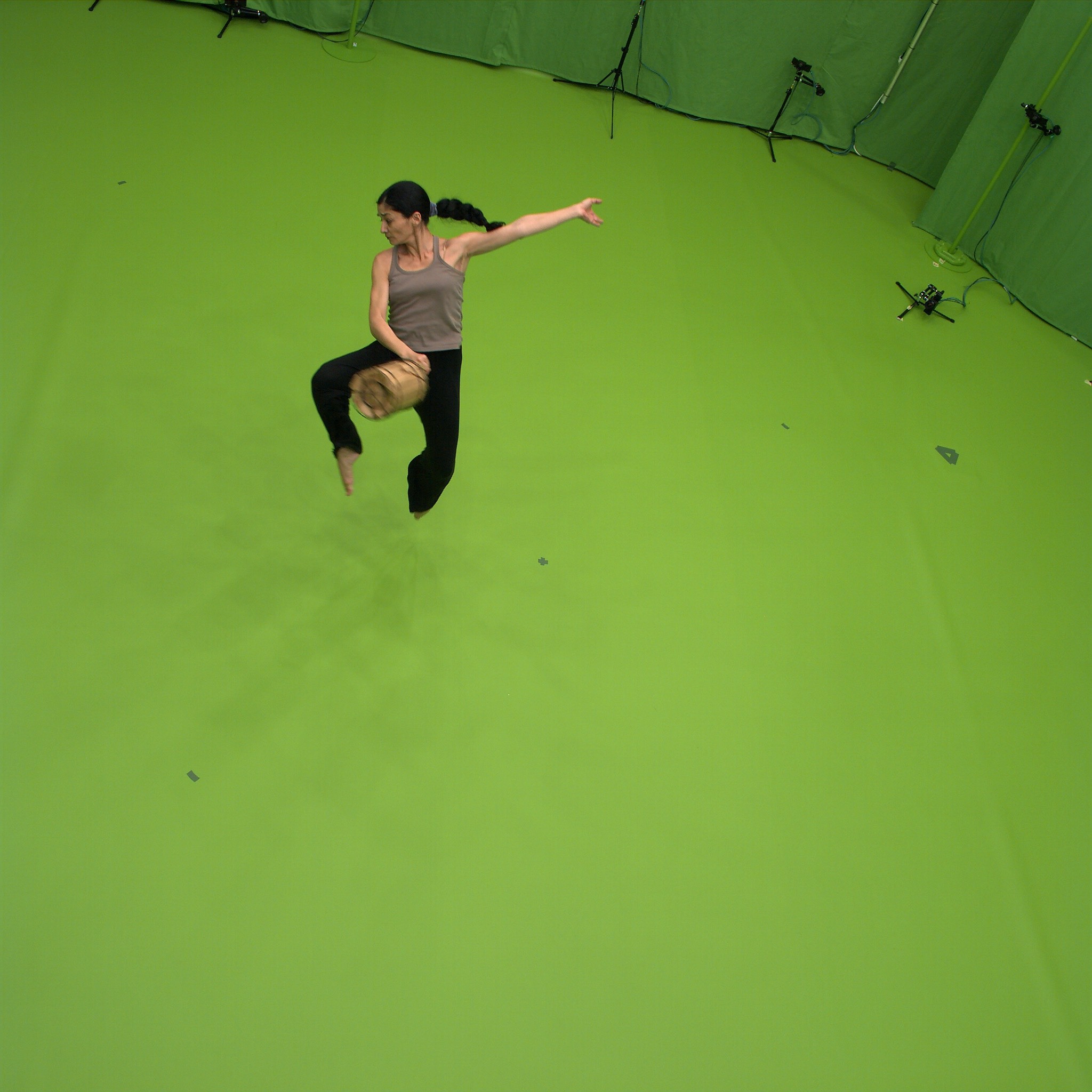} &
 	\includegraphics[width=.25\linewidth]{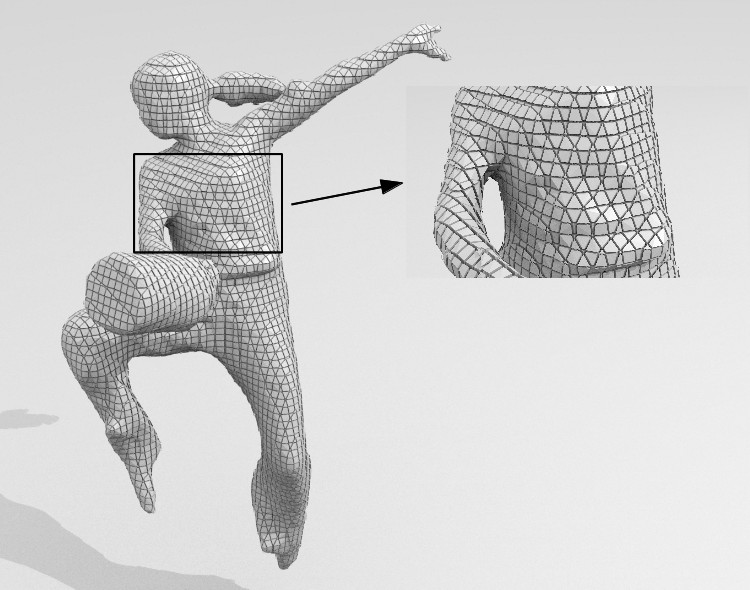} &
  	\includegraphics[width=.25\linewidth]{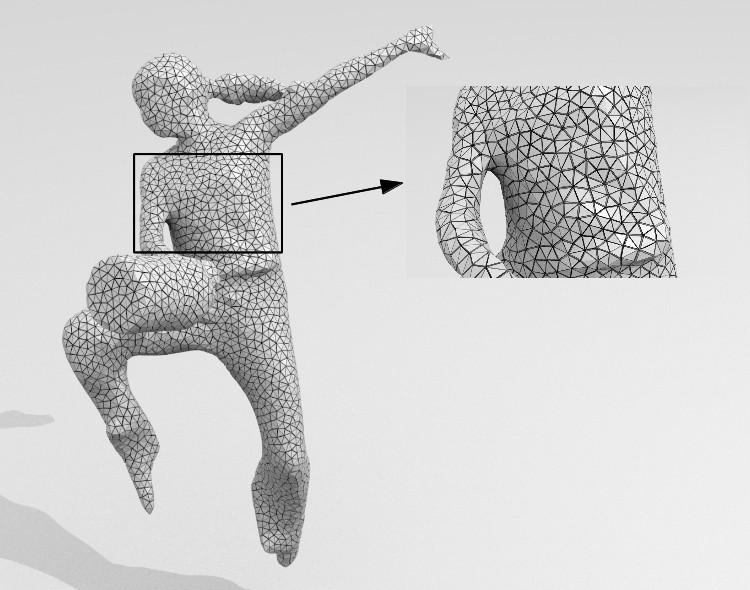} &
  	\includegraphics[width=.25\linewidth]{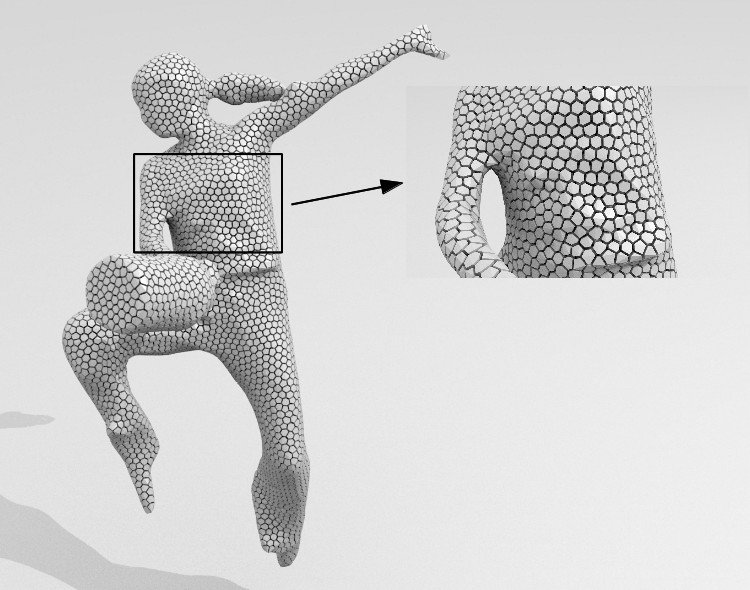} \\
  	(a) & (b) & (c) & (d)\\
  	\includegraphics[width=.13\linewidth]{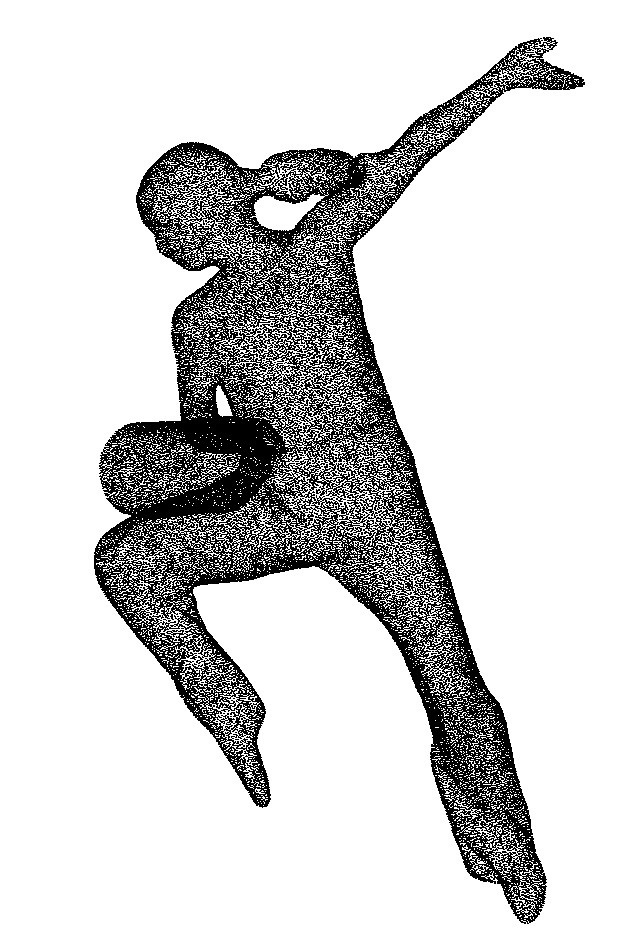} &
  	\includegraphics[width=.25\linewidth]{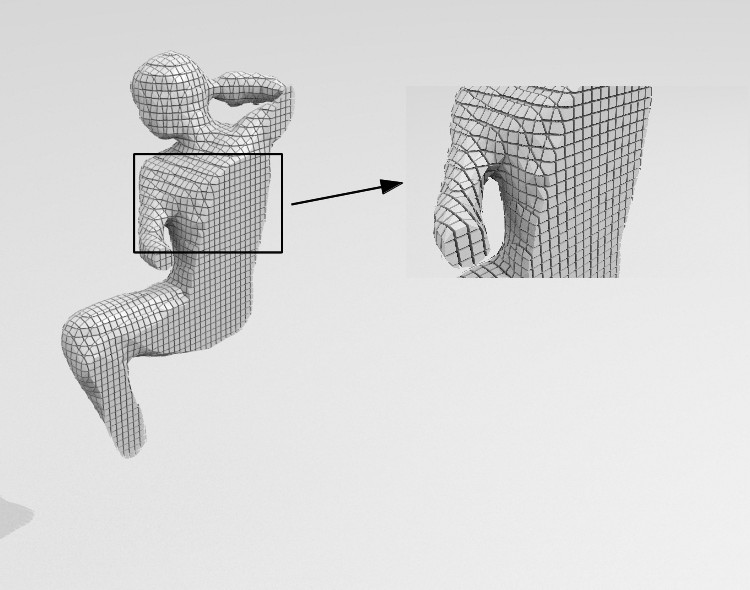} &
  	\includegraphics[width=.25\linewidth]{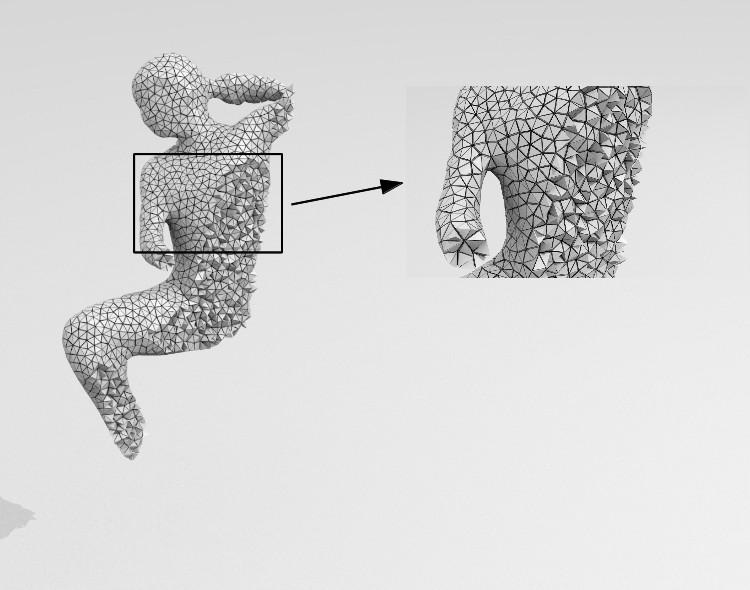} &
  	\includegraphics[width=.25\linewidth]{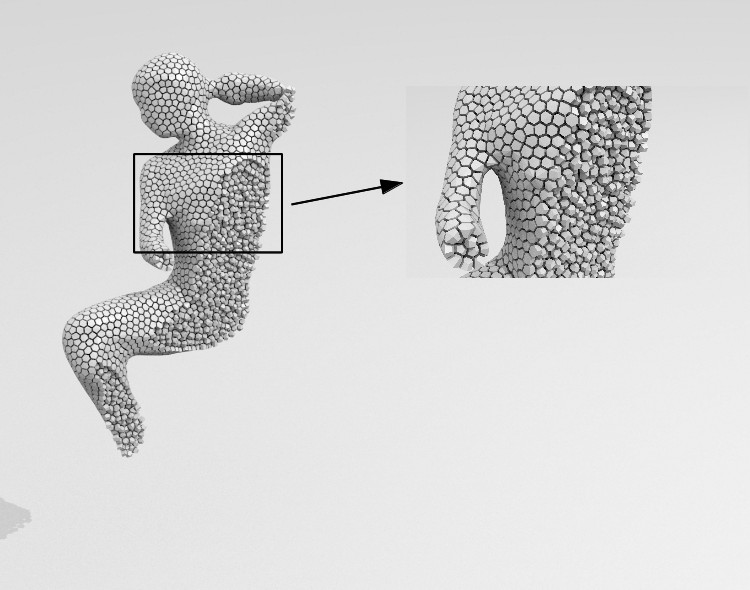} \\
  	(e) & (f) & (g) & (h)\\
  	\end{tabular}
\caption{(a, e)~Input multi-camera observation and point clouds (65386 pts).  (b,f)~Tessellations generated using voxels~\protect\cite{chernyaev95}. (c,g)~Tetrahedrisations generated using  Delaunay refinement~\protect\cite{cgal}  . (d,h)~Clipped Centroidal Voronoi Tessellations (14455 sites).} \label{fig:cvt2}
\end{figure*}

The algorithm was implemented in C++, and uses the libLBFGS library~\cite{lbfgs} for the L-BFGS computation during the optimization. The Tetgen library~\cite{tetgen} was used to compute Voronoi tessellations. Our approach was tested on oriented point clouds acquired with a multi-view  systems. Implicit functions were estimated from point clouds with the CGAL~\cite{cgal} implementation of the Poisson reconstruction algorithm. Figure~\ref{fig:cvt2} and Table~\ref{tab:cvt} show a comparison of  different strategies to get volumetric representations of a dancer.  This comparison was performed on various examples with similar results and we only present the dancer example for conciseness. The compared methods are a voxel method with a Marching Cubes  algorithm with topology guarantees~\cite{chernyaev95}  and the CGAL implementation of the Delaunay refinement approach~\cite{cgal}. The number of cells was made similar in all approaches by choosing the number of cubes (2cm$\times$2cm)  intersecting  or fully inside the shape as the number of sites for CVT ($14455$ in the example) and by making the cube diagonal   the length constraint  for Delaunay ball diameters in Delaunay refinement.  Figure~\ref{fig:cvt2} illustrates the benefit of CVTs for regularity. Note in particular the irregular boundary cells with the voxel representation.  In addition, Table~\ref{tab:cvt}  indicates a better precision for CVTs, where the precision evaluates the quality of the shape approximation by summing the Euclidian distances from the observed points to the generated shape surface.  Nevertheless, the table also shows the increased computation time  with CVTs, in particular when iterating over site locations. Optimized implementations for CVTs may anyway compensate partially  for this additional computation cost.

\begin{table}
\centering
	\begin{tabular}{|c|c|c|c|}
		\hline
		Object & Method & Error (m) & Time (s) \\
		\hline
%		Runner & MC & $9.529 \times 10^{-3}$ & 0.512 \\
%			   & Delaunay ref. & $22.740 \times 10^{-3}$ & 6.772 \\
%			   & CVT (0 iter.) & $9.372 \times 10^{-3}$ & 2.061 \\
%			   & CVT (10 iter.) & $6.262 \times 10^{-3}$ & 17.166 \\
%		\hline
		Dancer   & MC & $61.21$ & 0.945 \\
			65386 pts   & Delaunay & $147.31$ & 6.944 \\
			   & CVT (0 iter.) & $60.17$ & 2.14 \\
			   & CVT (10 iter.) & $42.77$ & 16.650 \\
		\hline
	\end{tabular}
\caption{Distance sum from the input point clouds (Figure~\ref{fig:cvt2}-(e)) to the estimated shape surface and computation times for:  Voxel representation (MC)~\protect\cite{chernyaev95}, Delaunay refinement strategy~\protect\cite{cgal} and our CVT approach.} 
\label{tab:cvt}
\end{table}

%-------------------------------------------------------------------------
\section{Volumetric Shape Tracking}
\secl{sec:tracking}

With the volumetric decomposition proposed above, we now need to define a model by which scene dynamics can be captured through deformations expressed over this decomposition. We consider here as input a time sequence of inconsistent CVT decompositions independently estimated at each frame and we look for a time consistent volumetric decomposition that encode cell motions. We opt for a capture by deformation approach where a template CVT, taken from the input sequence in our case, is tracked throughout the sequence.  This volumetric strategy    is motivated by two observations. First,  attaching the deformation model to the cell shape representation used for the animation directly provides the necessary cell dynamic information to the simulation. It avoids therefore the interpolation between an intermediate motion model, \eg a skeleton or a mesh, and the animation model; Such interpolation being  difficult to perform consistently over time. Second, as shown in~\cite{allain15}, it provides a simple tool for embedding volume-preservation constraints that increase the robustness of the tracking over the dynamic scenes we consider.  We describe below the generative approach~\cite{allain15} that we follow.

%intro: argumentation of usual models, kinematic skeletons and surface laplacian deformation.
%
%what are the inputs.
%
% more focus on patch based models. we improve and adapt patch-based models to a volumetric parametrerization.
% 
%
%following: after decsribing the motion parameterization, we provide a description of a probabilistic generative model of how our motion model explains measurements. then we explain how the model parameters can be inferred form measurements to complete the capture process.
\subsection{Tracking Formulation}

We are given a sequence of CVTs $\mathcal{V}$ and a template model $\hat{V}$. $\hat{V}$ can be one model  taken from the sequence or any other model (\eg a 3D scan) decomposed into a CVT. The tracking consists then in fitting $\hat{V}$ to each  $V \in \mathcal{V}$. This can be formulated as a maximum a posteriori (MAP) estimation of the deformation parameters $\hat{\Theta}$ that  maximizes the posterior distribution $\mathcal{P}(\Theta | \mathcal{V})$  of the parameters $\Theta$ given the observations $\mathcal{V}$: 
\[
\hat{\Theta} = \operatorname*{arg\,max}_{\Theta} ~ P(\Theta | \mathcal{V}) \simeq\operatorname*{arg\,max}_{\Theta} ~P(\mathcal{V}|\Theta) ~ P(\Theta).
\]
%where  $P(\mathcal{V}|\Theta)$ is the likelihood of the observations given the motion  parameters and  $\mathcal{P}(\Theta)$ is the prior information on these parameters. 
Taking the log of the above expression yields the following optimization problem:
\begin{equation}
\hat{\Theta} = \operatorname*{arg\,max}_{\Theta} ~ E_{data}(\mathcal{V}, \Theta) + E_{prior}(\Theta),
\label{dgr:eq:opt}
\end{equation}
where the data term $E_{data}$ evaluates  the log-likelihood of a set of deformation parameters $\Theta$ given the observations $\mathcal{V}$, and the regularization term $E_{prior}$ enforces prior constraints on the deformation, \eg local rigidity.  We detail below the parameterization $\Theta$ we use for the deformation and the associated energy terms.

\subsection{Motion Parameterization}

\begin{figure}[tpb]
\centering
%\begin{tabular} {cc}
\frame{\includegraphics[width=.50\linewidth]{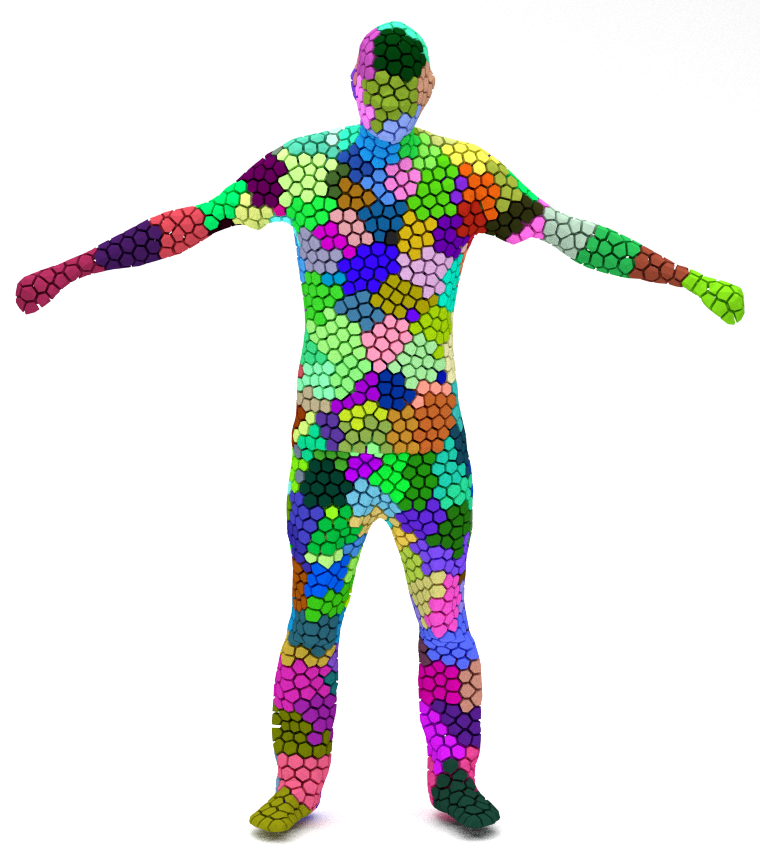}}
%\end{tabular}
\caption{The template model used to recover  the runner sequence motion, with its CVT decomposition cells, and the cell clusters in different colors.}
\figl{clusters}
\end{figure}

The deformation model is defined over CVT cells in the shape decomposition and, for efficiency, on aggregates of cells which reduces the number of terms and parameters. To this goal, CVT cells are grouped together as a set of volumetric patches \patch using a k-medoids algorithm, as shown in Figure~\ref{fig:clusters}. Such patches can be either adjacent to the surface or completely inside the template shape's volume, which is of particular interest to express non-rigid deformation of the model while preserving the local volume and averting over-compression or dilation. The positional information of a patch is represented as a rigid transform $\patchpose \in SE(3)$ at every time~\tidx.
Each position \x*[k,\vidx] of a CVT sample is indiscriminately labeled as a point~\vidx. Its position can be written as a transformed version of its template position \xref as follows, once the patch's rigid transform is applied:

\begin{align}
  \x*[k,\vidx] = \patchpose*[k] (\xref).
  \eql{verttrans}
\end{align}

A \textit{pose} of the shape is thus defined as the set of patch transforms $\patchpose*= \set{\patchpose*[k]}[k \in \mathcal{K}]$, which expresses the deformation of every component in the shape. The parameterization $\Theta$ is then the set of pose parameters  of the template over the considered time sequence $\mathcal{T}$:
\[
\Theta = \{\patchpose*^t\}_{t \in \mathcal{T}}.
\]

\subsection{Data Term}
\secl{observations}

We assume the observed shape $V^{\tidx}$ at time $t$ is described by the point cloud $\Obs<t> = \{\obs\}$. We assume this point cloud to include inner volume points and surface points, \ie the CVT sites as well as the outside surface points.

 In order to measure how a deformed version of the template explains the observed shape, first the associations between the observations and the template must be determined. This is achieved via a soft ICP strategy that iteratively reassigns each observation \obs to  the template volumetric patches.  For simplicity, each observation \obs is associated to the patch \patch via the best candidate point \x*[k,\vidx] of patch \patch and with an association penalty $\alpha_{\symb{o}, k}$. 
 %The best candidate is chosen as the closest compatible point in the patch \patch during iterative resolution. The association penalty $\alpha_{\symb{o}, k}^t$ accounts for the distance between \obs and the best candidate of \patch at a previous iteration. 
 
 The matching penalty $E( \obs, \patchpose[\ksel*] )$ that evaluates how well \patch explains \obs is then the weighted distance between \obs and the best candidate \x*[k,\vidx], that is  the template point \xref[\vidx] transformed by  the current pose \patchpose[\ksel*] of the template model: 
\begin{align}
	\eql{obslikelihood}
  	E( \obs, \patchpose[\ksel*]) \; &= \; \alpha_{\symb{o}, k}^t  \; \|\obs - \patchpose[\ksel*](\xref[\vidx])\|.
\end{align}
Associations are additionally filtered using a compatibility test. Observed surface points are associated to template surface points  with similar orientations with respect to a user defined threshold $\theta_{max}$; Observed inner  points are associated to template inner points that present  similar distances to the surface up to a user defined tolerance $\epsilon$.  If there is no compatible candidate in a patch \patch, then \patch is discarded for the association with \obs, \ie  $\alpha_{\symb{o}, k}^t = 0$ . Finally:
\begin{equation}
E_{data}(V^t, \Theta^t)= \sum_{\symb{o}, k} \;  E( \obs, \patchpose[\ksel*]). 
\label{eq:dataterm}
\end{equation}

\subsection{Regularization Term}
The pose of a shape is defined by the set of rigid motion parameters of the shape volumetric patches. While these parameters hardly constraint the patch motions, they do not define a coherent shape motion since each patch moves independently of the others. In order to enforce  shape cohesion, soft local rigidity constraints, reflecting additional prior knowledge on shape deformation parameters, are considered. These constraints rely on a pose distance function that evaluates how distant from a rigid transformation a deformation between two poses is. 
Once such a distance is defined, the regularization term is defined as the sum of distances between all poses in the sequence and a given pose that can be a reference pose or an estimated mean pose, as explained below.

\paragraph*{Pose Distance} To simplify the estimation, the shape distance is expressed over coordinates of points belonging to  patches and not on the parameters of the pose itself:
\begin{align}
\eql{energy}
 \distance(\patchpose*<i>,\patchpose*<j>) &= \sum_{(\patch,\patch[l]) \in \mathcal{N}}  \distance[kl](\patchpose*<i>,\patchpose*<j>), \quad \mbox{with}  \\
\distance[kl]( \patchpose*<i>,\patchpose*<j>) &=\sum_{\vidx \in \patch[k] \cup \patch[l]} \|\patchpose[k-l]<i>(\xref) - \patchpose[k-l]<j>(\xref)\|^2, \nonumber
\end{align}
where $\patchpose[k-l]<i>={\patchpose[l]<i>}^{-1} \circ \patchpose[k]<i>$ is the relative transformation between patches \patch and \patch[l] for pose $i$, and $\mathcal{N}$ is the set of neighboring patch pairs within the shape.  Intuitively, this distance measures whether the relatives poses between neighboring volumetric patches are preserved during motion between two shape poses. 

%The distance would be zero if all patches undergo  single rigid motions in both poses: $\patchpose[k-l]<i>=\patchpose[k-l]<j>=0$ or if all patches undergo different rigid motions but anyway equal for patches in both poses and up to a global rigid motion per pose. 

\paragraph*{Deformation Energy}
\secl{deformation}

The pose distance above allows to compute a deformation energy between two poses and with respect to a reference pose from which the patch transformations $\patchpose[t]<k>$ are expressed. This reference pose can be taken as the identity pose of the initial template model (see Figure~\ref{fig:clusters} for instance). However such a strategy is biased toward the template pose and discourage locally rigid motions between poses that are distant from the template pose. A more appropriate approach is to exploit the pose distance function to first define a mean pose  $ \patchposem*$ over a time window $\{t\}$: 
\[
\patchposem* =\operatorname*{arg\,min}_{\mathbf{T}}  \sum_{t} \distance(\patchpose*<\tidx>, \mathbf{T}).
\]
 This averaged pose can then be taken as an evolving reference pose from which  non-rigid deformations are measured to define  the deformation energy over the associated time interval: 
\begin{align}
 \eql{poselikelihood}
 \E(\{\patchpose*<\tidx>\}, \patchposem*)\; &= \; \sum_{\tidx } \distance(\patchpose*<\tidx>, \patchposem*).
\end{align}
This imposes general proximity of poses to a sequence specific ``rest'' pose. These rest poses must also be constrained to some form of inner cohesion. This is ensured  by minimizing the distance from the mean pose to the identity pose of our initial template:
\begin{align}
\eql{restprior}
 \E( \patchposem* ) \; = \;  \distance(\patchposem*,\mathbf{Id}),
\end{align}
%Because the template pose defines a neutral set of volumetric patch positions, the latter ensures that the mean pose \patchposem* of any sequence has inner relative patch transforms similar to the template initial relative patch poses, preserving the geometric features of the template. This preservation is in turn transferred to all poses in the sequence thanks to the constraint \eqn{poselikelihood}. 
This definition of deformation has a number of advantages: first it enforces geometric cohesion and feature preservation, and second it is quite simple to formulate and to optimize, as the minimization of  \eqn{poselikelihood} and \eqn{restprior} translates to a sum of least square constraints over the set of CVT sites.The prior energy term finally writes:
\begin{align}
\eql{prior}
 E_{prior}(\Theta=  \{\patchpose*^t\}) =   \;  \distance(\patchposem*,\mathbf{Id})  \;+  \; \E(\{\patchpose*<\tidx>\}. \patchposem*),
\end{align}
%Note that if we replace in the above expression the mean pose by the template pose, we get the energy of the strategy mentioned earlier with a stronger influence of the template pose. 

We jointly extract poses and a sequence mean pose by minimizing the sum of the data terms (\ref{eq:dataterm}) and the prior term \eqn{prior}. Section \secref{results} shows results on various sequences and gives run time performances.

%\subsection{Optimization}
%\secl{inference}
%%se referre au papier eccv 2014 etendu au framework
% 
%%\subsection{Joint Distribution}
%
%We jointly extract poses and a sequence mean pose by minimizing the sum of the data term \eqn{obslikelihood} and prior term \eqn{prior} given observations $\{\oidx\}$ over a time window $\{t\}$ and a template model with patches  $\{\patch\}$:
%
%
%
%\begin{align}
%\eql{joint}
%\argmin*[\patchposem*, \{\patchpose*<\tidx>\}] \; \E(\patchposem*) +\E(\{\patchpose*<\tidx>\}, \patchposem*) + \sum_{\oidx, t, k} E( \obs, \patchpose ) 
%\end{align}
%% 
%It can be shown that this yields a set of least squares constraints over mean pose \patchposem* and sequence poses $\{\patchpose*<\tidx>\}$, which can be efficiently solved with standard tools. In ICP fashion, we alternate this minimization with the association step between observations and patches. Section \secref{results} shows results on various sequences and gives run time performances.  
%In practice, we observe that using averaging over several candidates for a given observation \oidx yields more robust results and we use this in our implementation.

%-------------------------------------------------------------------------
\section{Combined Animation}
\secl{sec:anim}

\newsymbol{\state}{\mathbf{X}}(t)
\newsymbol{\dstate}{\dot{\mathbf{X}}}(t)
\newsymbol{\cm}{\hat{x}}(t)
\newsymbol{\dcm}{\dot{\hat{x}}}(t)
\newsymbol{\rot}{R}(t)
\newsymbol{\drot}{\dot{R}}(t)
\newsymbol{\v}{v}(t)
\newsymbol{\dv}{\dot{v}}(t)
\newsymbol{\P}{P}(t)
\newsymbol{\dP}{\dot{P}}(t)
\newsymbol{\L}{L}(t)
\newsymbol{\dL}{\dot{L}}(t)
\newsymbol{\I}{I}(t)
\newsymbol{\w}{\omega}(t)
\newsymbol{\M}{M}
\newsymbol{\F}{F	}(t)
\newsymbol{\torque}{\tau}(t)

The template representation of the subject now being consistently tracked across the sequence, we can use the tracked cells as input for solid physics-based animation.
We have purposely chosen CVTs as a common representation as they can be made suitable for shape and motion capture as we have shown, while being straightforwardly convenient for physics-based computations. In fact CVT cells are compact, convex or easily approximated by their convex hull. This is an advantage for the necessary collision detection phase of physics models, as specific and efficient algorithms exist for this case~\cite{gilbert88,rabbitz94}. We here describe the common principles of our animation model, with more specific applications being explored and reported on in the following sections.

As our animation framework is solid-based, we base our description on commonly available solid-based physics models, \eg \cite{baraff97}. Each CVT cell is considered a homogeneous rigid body, whose \emph{simulated} state is parameterized by its 3D position, rotation, linear momentum and angular momentum.
The cell motion is determined by Newton's laws through a differential equation involving the cell state, the sum of forces and sum of torques that are applied to the cell.

The animation is thus obtained by defining the set of forces and torques applied at each instant, and iteratively solving these differential equations to obtain a new cell position and orientation for a target time step, using one of many available techniques. For our demonstrator, we use the simple and efficient off-the-shelf Bullet Physics engine~\cite{bullet,catto05}.

\subsection{Ordinary Applied Forces}

We classically apply the constant gravity force $\F*[g] = \M g$. We apply additional external forces or constraints as needed for the target application, as will be detailed in the coming sections. Additionally, contact forces such as collisions are handled with scene objects, as well as between different cells of the CVT. 
% To avoid over-constraining the model, whose CVT cells are densely packed in the simulation,   .
For this purpose the physics engine first needs to detect the existence of such contacts. It relies on a hierarchical space decomposition structure, such as an AABB-tree, for broad-phase collision detection, \ie coarse elimination of collision possibilities. A narrow-phase collision test follows, between objects lying in the same region of space as determined by the AABB-tree traversal. In this narrow phase the full geometry of objects is examined, \eg using the GJK algorithm~\cite{gilbert88} for pairs of convex polyhedra. Once the existence of a contact is established, various strategies exist to deal with the collision, \eg by introducing impulse repulsion forces to produce a collision rebound. %~\cite{}.
We follow the common approach of modeling contacts as a linear complementary problem (LCP) popularized by~\cite{baraff94}, which derives contact forces as the solution of a linear system that satisfies certain inequality constraints. These constraints are typically formulated using a constraint Jacobian over the combined state spaces of rigid bodies. \cite{catto05} expose the specific variant applied in the context of the Bullet Physics engine.

\subsection{Physical Modeling of Kinematic Control}

To relate the physical simulation to the acquired non-rigid poses of the model, we need to introduce coupling constraints. Our goal is to allow the model to materialize and control the tradeoff between the purely kinematic behavior acquired from visual inputs for the cell, and the purely mechanically induced behavior in the simulation. First it is important to note that the temporal discretization used for acquisition and for simulation and rendering of the effects are generally different. Consequently the first stage in achieving our goal is to compute a re-sampling of the pose sequence, to the target simulation and rendering frequency, using position and quaternion interpolation. The poses so obtained are here referred as the acquired cell poses \cm<a> and \rot<a>. Second, we formulate the coupling by introducing a new kinematic recall force, in the form of a damped spring between the acquired cell poses and the simulated cell poses:
\begin{equation}\label{eq:recallforce}
\F[r] = k.(\cm<a> - \cm) - \lambda. \frac{\textrm{d}}{\textrm{d}t}(\cm<a> - \cm),
\end{equation}
where $k$ and $\lambda$ are respectively the rigidity and damping coefficients of the spring, which control the strength and numerical stability of the coupling.

%-------------------------------------------------------------------------
%\section{Results}
\section{Visual Effects}
\secl{results}

\begin{figure*}[htpb]
%\centering
\hspace*{-4mm}
\centering
\begin{tabular} {c}
\includegraphics[width=0.95\linewidth]{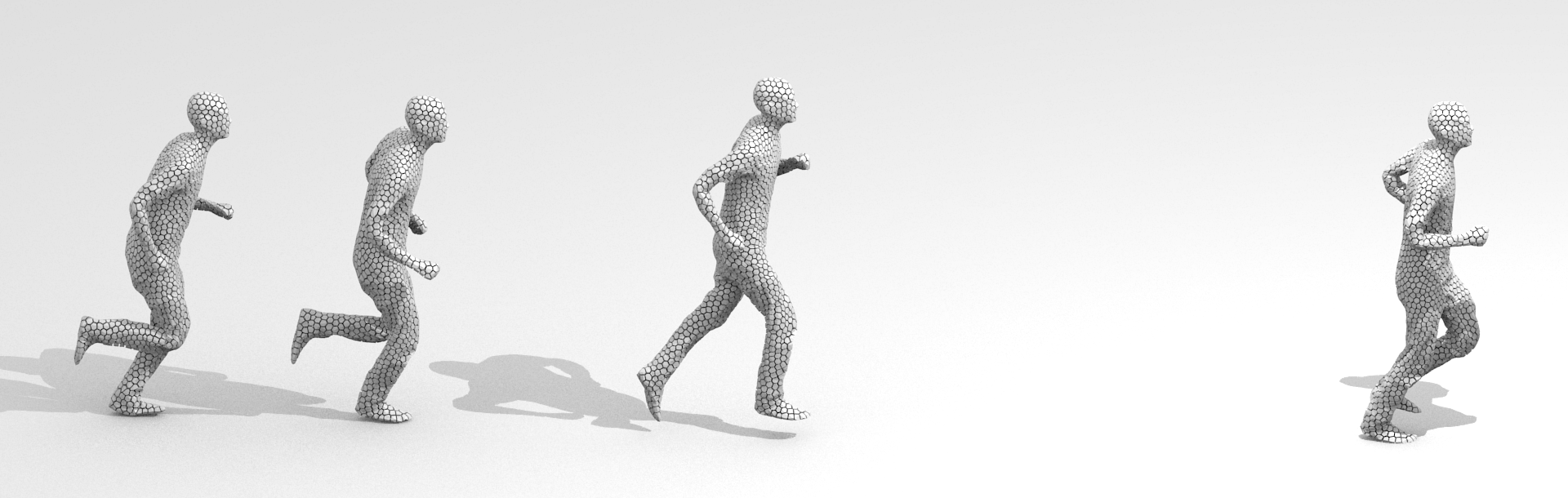}\\
\includegraphics[width=0.95\linewidth]{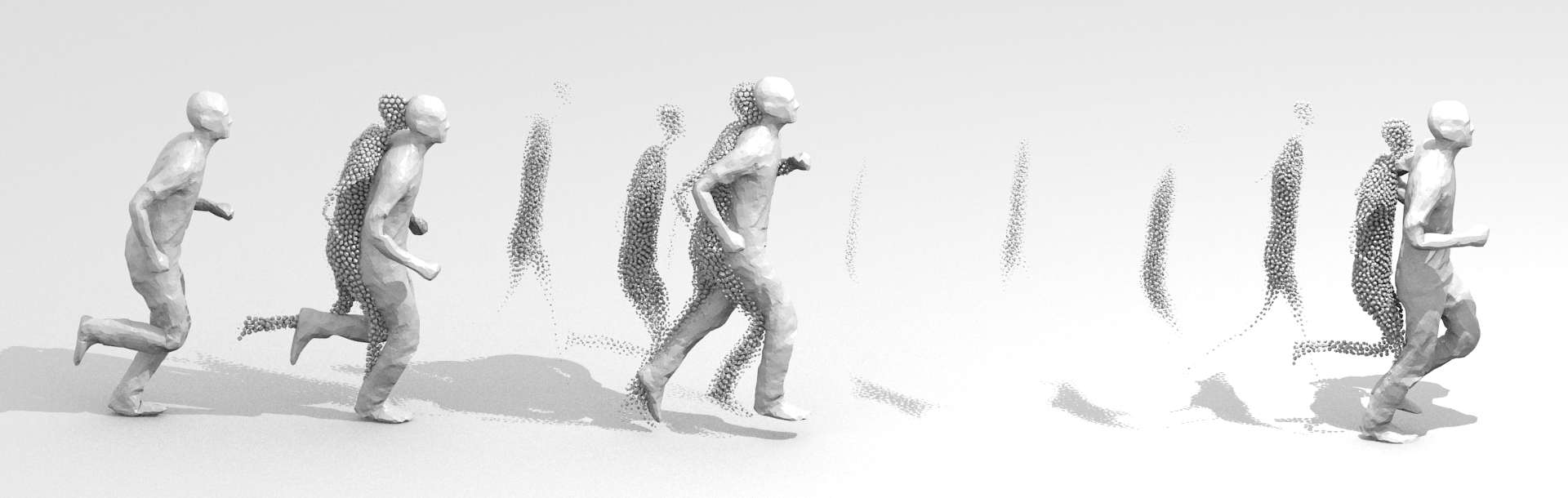}
\end{tabular}
\caption{Time persistence on the \runner{} sequence: a slower copy of the shape that erodes over time is generated at regular intervals. }
\figl{persistence}
\end{figure*}

This section presents various animation results on three captured animations of variable nature and speed (see also the accompanying video). \runner{} %(\fig{fig:teaser} and \fig{fig:runner-slackline}, left)
shows a male character running in a straight line, during 3 motion cycles. This animation lasts 2.5 $s$. In \cagebirddance{} %(\fig{fig:dance}),
a female dancer moves while holding a bird cage. This sequence is 56 $s$ long and shows a complex sequence of motions which would be difficult to synthetize without sensors. Finally, in \slackline{} %(\fig{fig:runner-slackline}, right)
a male acrobat evolves on a non rigid line above the ground, for 25 $s$.
%Our setup includes 68 cameras at 50 fps.
The input sequences of temporally inconsistent 3D point clouds are made respectively of 126 (\runner{}), 2800 (\cagebirddance{}) and 1240 (\slackline{}) temporal frames. 

\paragraph*{Parameters} The interior of all shapes has been tessellated according to \secref{sec:cvt} using 5000 cells. 10 iterations of the L-BFGS quasi-Newton algorithm were applied, except for the template shape where 50 iterations were applied. We use a temporal window of 10 frames for the tracking, and cluster the 5000 cells into 200 patches. 60 iterations were applied.

%Graphical results: ie. 2-3 sequences with different effect, material, etc.
We list below examples of visual effects that we were able to generate using the proposed approach. %TODO: attention, dans l'abstract on liste : distortion, erosion, morphing, gravity pull, compression => pb de cohérence !
First, we present animations combining tracking results with solid dynamics simulation (\secref{kinematicdeactivation}). Then we present other visual effects that also exploit the volumetric tracking information (\secref{persistence} and \secref{morphing}).
\subsection{Asynchronous Kinematic Control Deactivation}
\secl{kinematicdeactivation}
In order to show the effect of gravity while keeping the dynamic motion of the input sequence, we deactivate kinematic control forces independently for each cell. After being deactivated, a cell usually falls to the ground, since it follows a trajectory determined only by gravity, collision forces and its initial velocity. Asynchronous cell deactivations result in an animation combining cells that follow their tracking trajectory and cells that fall. By choosing diverse strategies for scheduling cell deactivations, a wide variety of animations can be obtained.

We explore here three possible deactivation strategies that are based on different criteria. Note that while these effects are straight-forward to produce with our volumetric framework, it would be difficult to obtain them if only surface or skeleton-based tracking was available.

\subsubsection{Rupture under Collisions}
Collisions with obstacles sometimes deviate a cell from its theoretical trajectory, which results in an increase of the recall force magnitude (see Eq.~\ref{eq:recallforce}). This phenomenon can be detected and used for simulating the rupture of the material: when the recall force magnitude of a cell is above a given threshold, we deactivate the recall force (for this cell only). This makes the rupture looks like the consequence of the collision.%, which results in a plausible animation.

Figure~\ref{fig:pendulum} shows a heavy pendulum that hits the subject and makes a hole in it. Under the intensity of the collision, several cells are ejected (rupture) and fall to the ground.

\subsubsection{Heat Diffusion}
In order to make the cell deactivation both temporally and spatially progressive, we rely on a diffusion algorithm. We diffuse an initial temperature distribution inside the volume according to the diffusion equation. Deactivation is triggered when the cell temperature is above a given threshold.

\paragraph*{Heat diffusion in a CVT}
The CVT provides a graph structure on centroids, which is a subgraph of the Delaunay tetrahedralization of the cells centroids.
The heat diffusion on a graph structure is expressed by the heat equation:
\[
( \partial / \partial {t} + \laplacian ) \heat = 0
\]
where $\heat$ is the column vector of centroids temperatures at time $t$, and $\laplacian$ is the combinatorial graph Laplacian matrix (note that a geometric Laplacian is not necessary since centroids are regularly distributed in space).
Given an initial temperature distribution \heat[0], this equation has a unique solution
\[
\heat = \heatkernel \heat[0]
\]
where $\heatkernel = e^{t\laplacian}$ is the heat diffusion kernel, which can be computed by means of the spectral decomposition of \laplacian .
%% Let ${(\eigenval, \eigenvect)}_{i=1}^{n}$ be the eigen values and eigen vectors of a spectral decomposition of the graph laplacian matrix $L$ ($\laplacian$ is symmetric). The heat diffusion at time $t$.
%% \begin{align}
%% \heat &= \heatkernel \heat[0]\\
%%\heatkernel &= \sum_{k = 2}^{n} e^{-t\eigenval} \eigenvect {\eigenvect}^{\top}
%%\end{align}

%\paragraph*{Experiments}
We compute the temperature evolution on the \cagebirddance{} sequence for an initial temperature distribution where all cell temperatures are zero, except for the dancer's head top cells (which are set to $1$). We observe in Figure~\ref{fig:teaser} that cells fall progressively across time, from the upper to the lower body parts. Note that the cage cells remain kinematically controlled since heat is not transferred between different connected components.

\subsubsection{Morphological Erosion}
The discrete cell decomposition of shapes allows to apply morphological operators. To illustrate this principle, we have experimented erosion as shown in Figure~\ref{fig:outline} .  In this example, cells are progressively eroded starting from the outside. The morphological erosion is performed by deactivating each cell after a delay proportional to the distance between the cell centroid and the subject's surface. The distance is computed only once for each cell, on the template shape. Figure~\ref{fig:outline} shows the erosion animation on the \runner{} sequence. Note that the operation progressively reveals the dynamic of the inner part of the shape. 

\subsection{Time Persistence}
\secl{persistence}
In this example,  we experiment time effects over  cell decompositions. To this aim,  dynamic copies of the model are generated at regular time intervals. These copies are equipped with  deceleration and erosion effects over time and create therefore ghost avatars that vanish with time (see Figure~\ref{fig:persistence} and the accompanying video). The benefit of  the tracked volumetric representation in this simulation  is the ability to attach time effect to the model behavior at the cell level, for instance lifetime and deceleration in the example.

\subsection{Morphing}
\secl{morphing}
\begin{figure*}[htbp]
\centering
\hspace*{-4mm}
\begin{tabular} {c@{\hskip 0.6mm}c@{\hskip 0.6mm}c@{\hskip 0.6mm}c@{\hskip 0.6mm}c@{\hskip 0.6mm}c}
\includegraphics[width=0.16\linewidth]{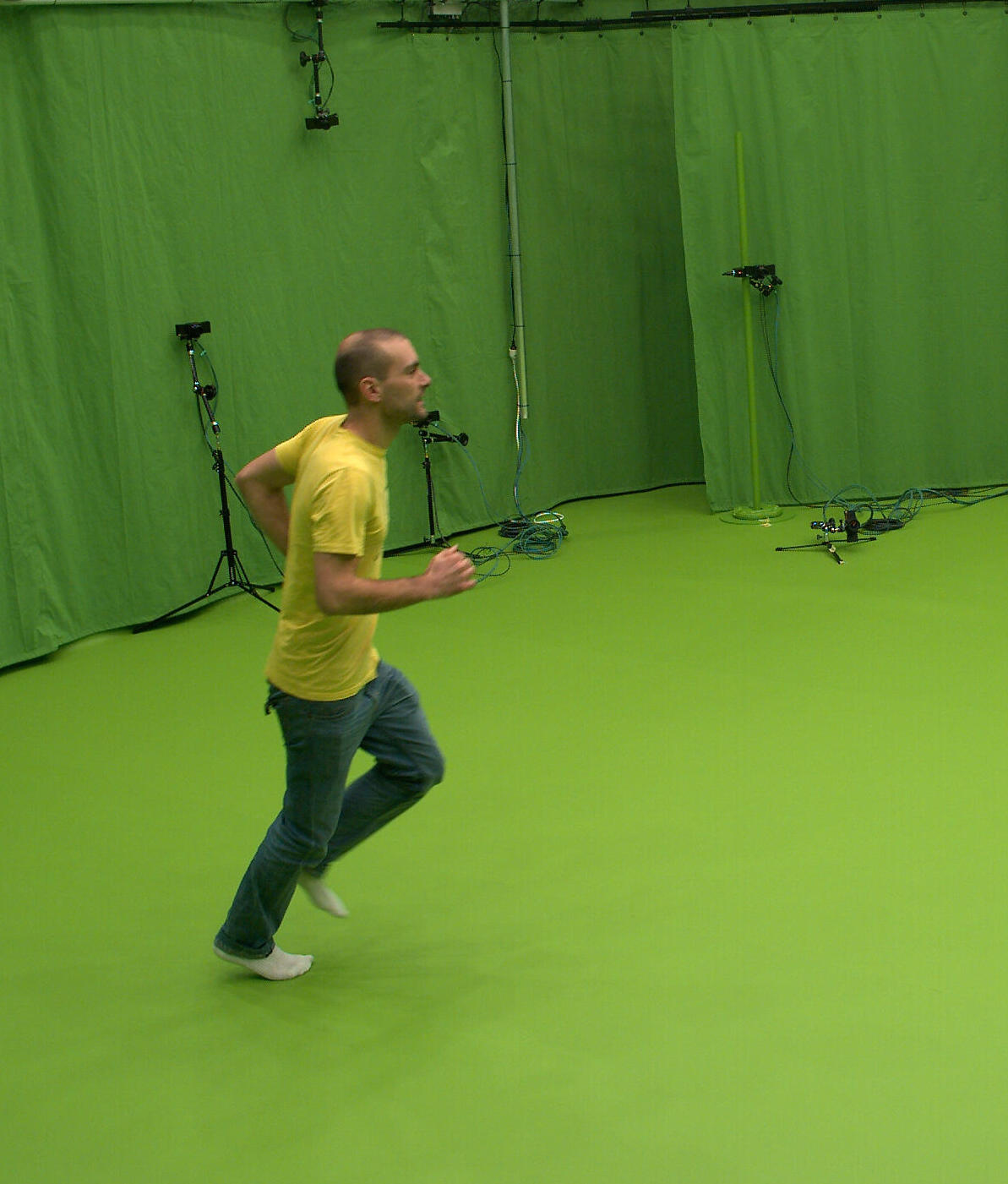} & \includegraphics[width=0.16\linewidth]{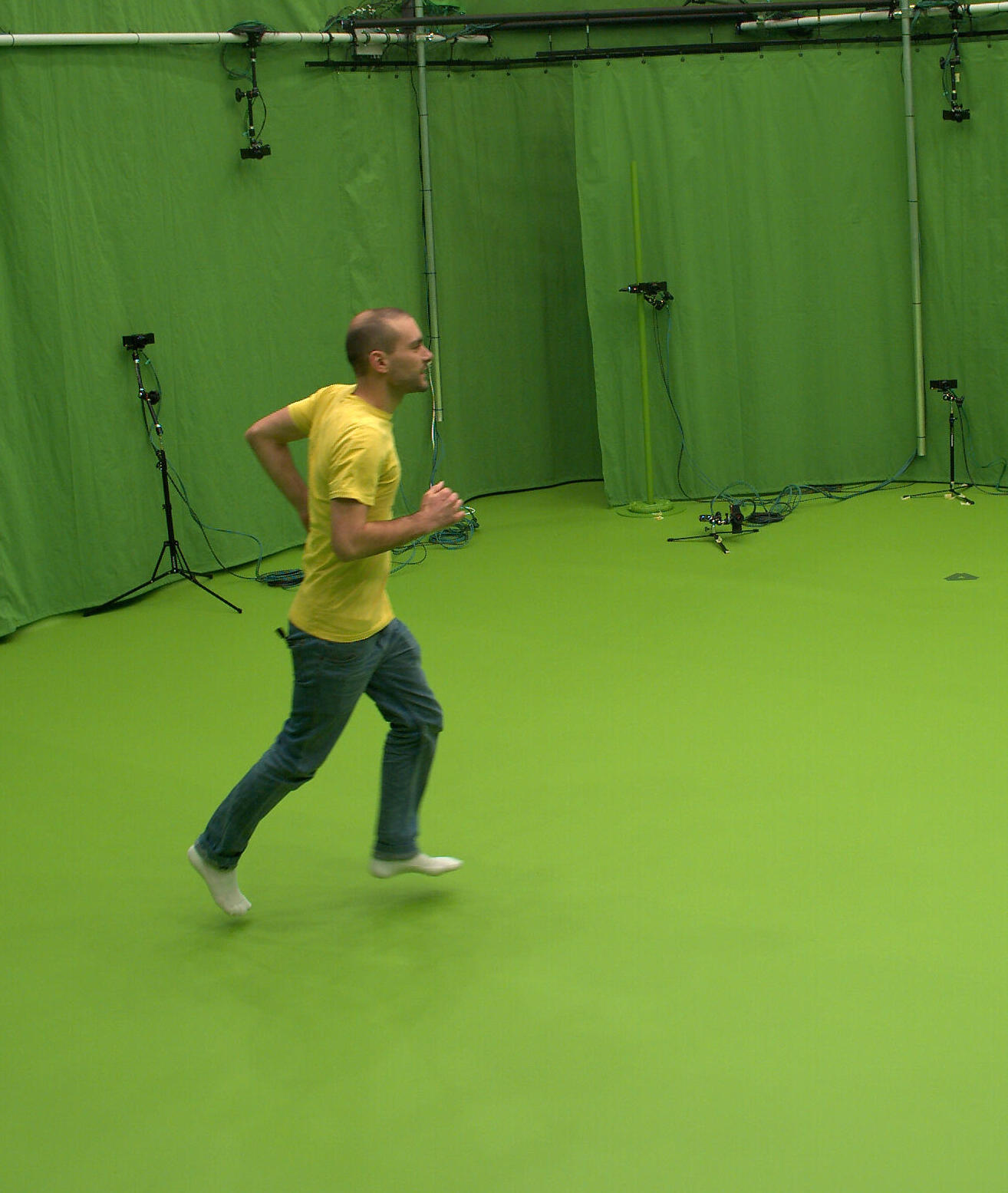} & \includegraphics[width=0.16\linewidth]{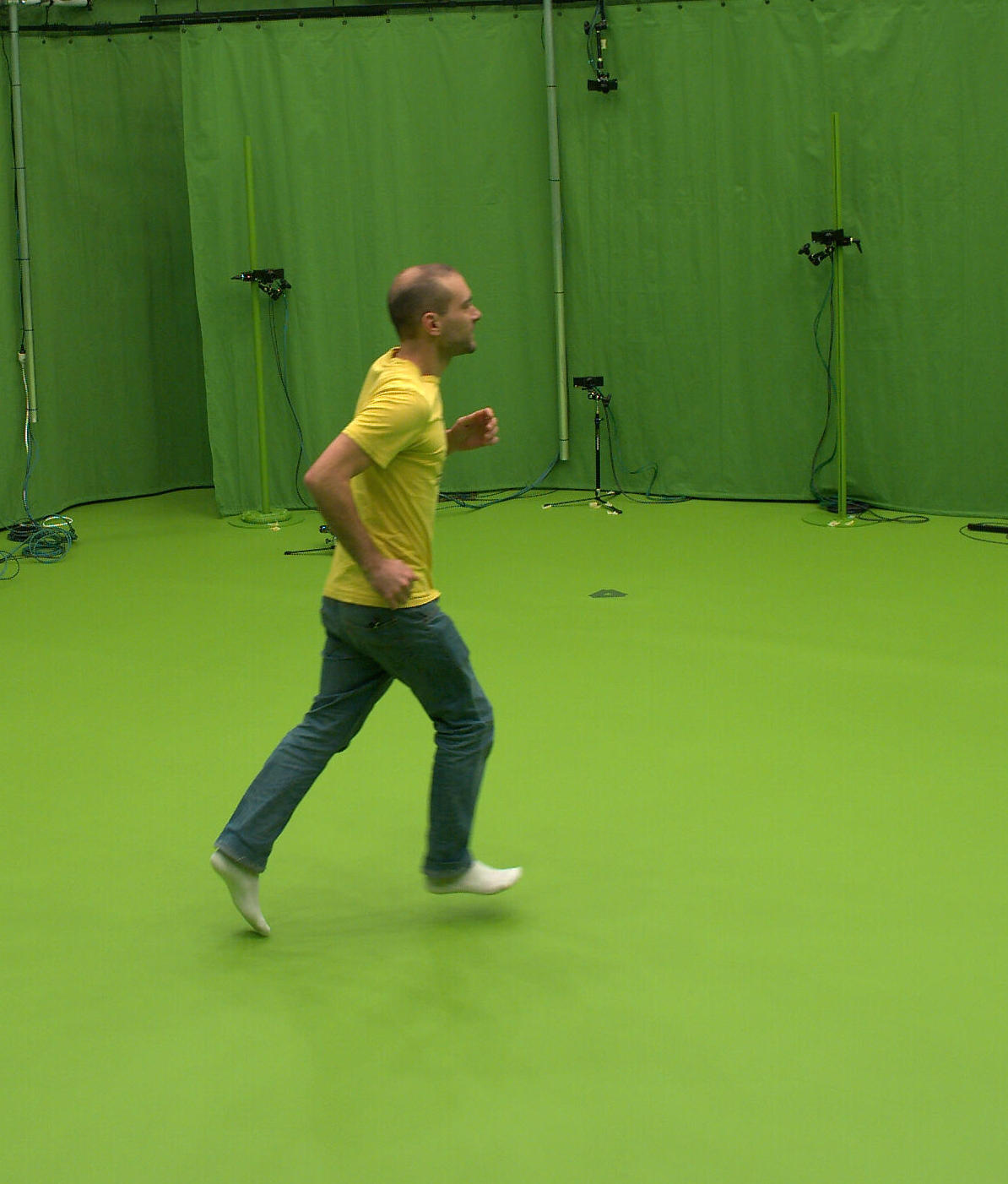} & \includegraphics[width=0.16\linewidth]{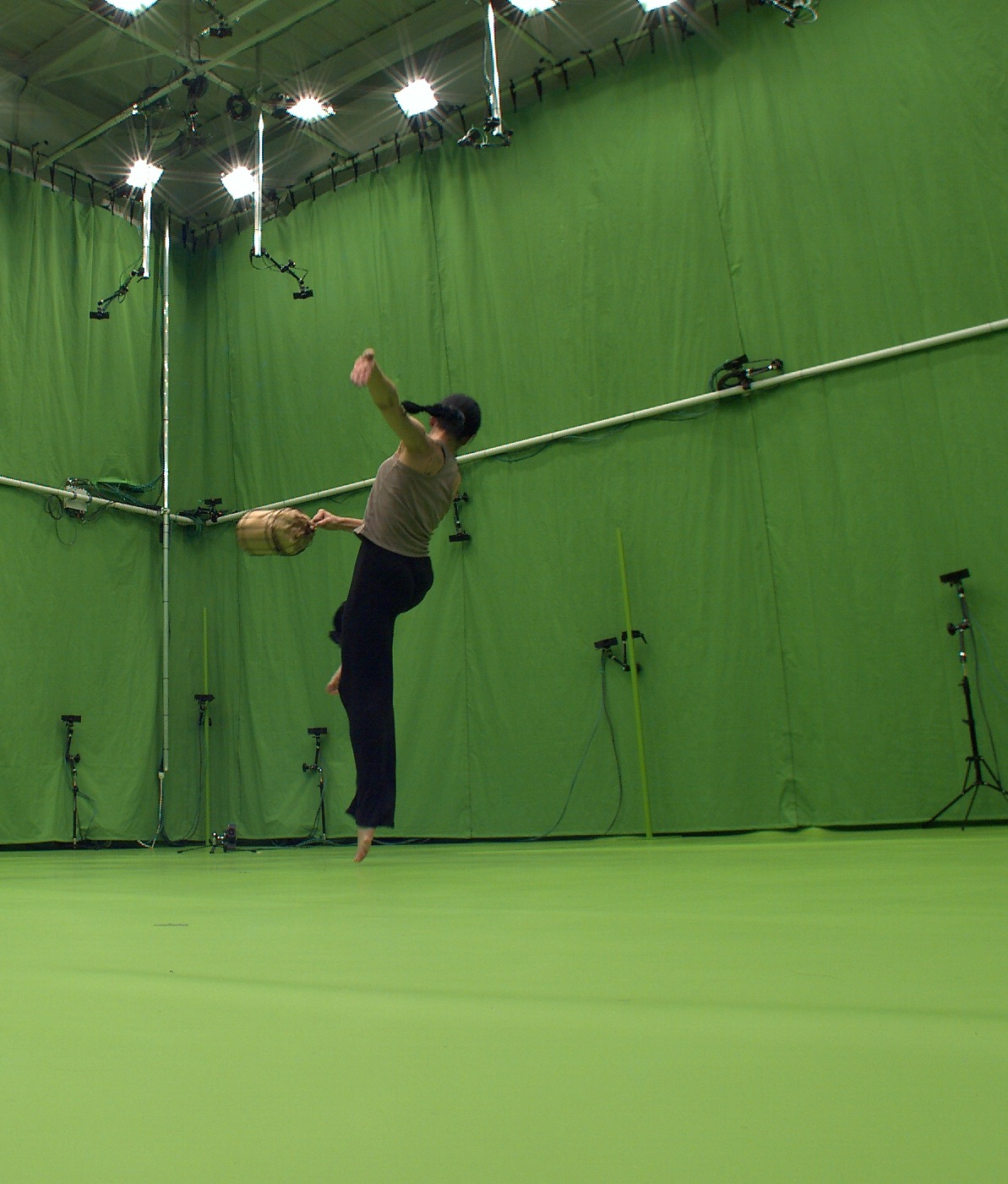} & \includegraphics[width=0.16\linewidth]{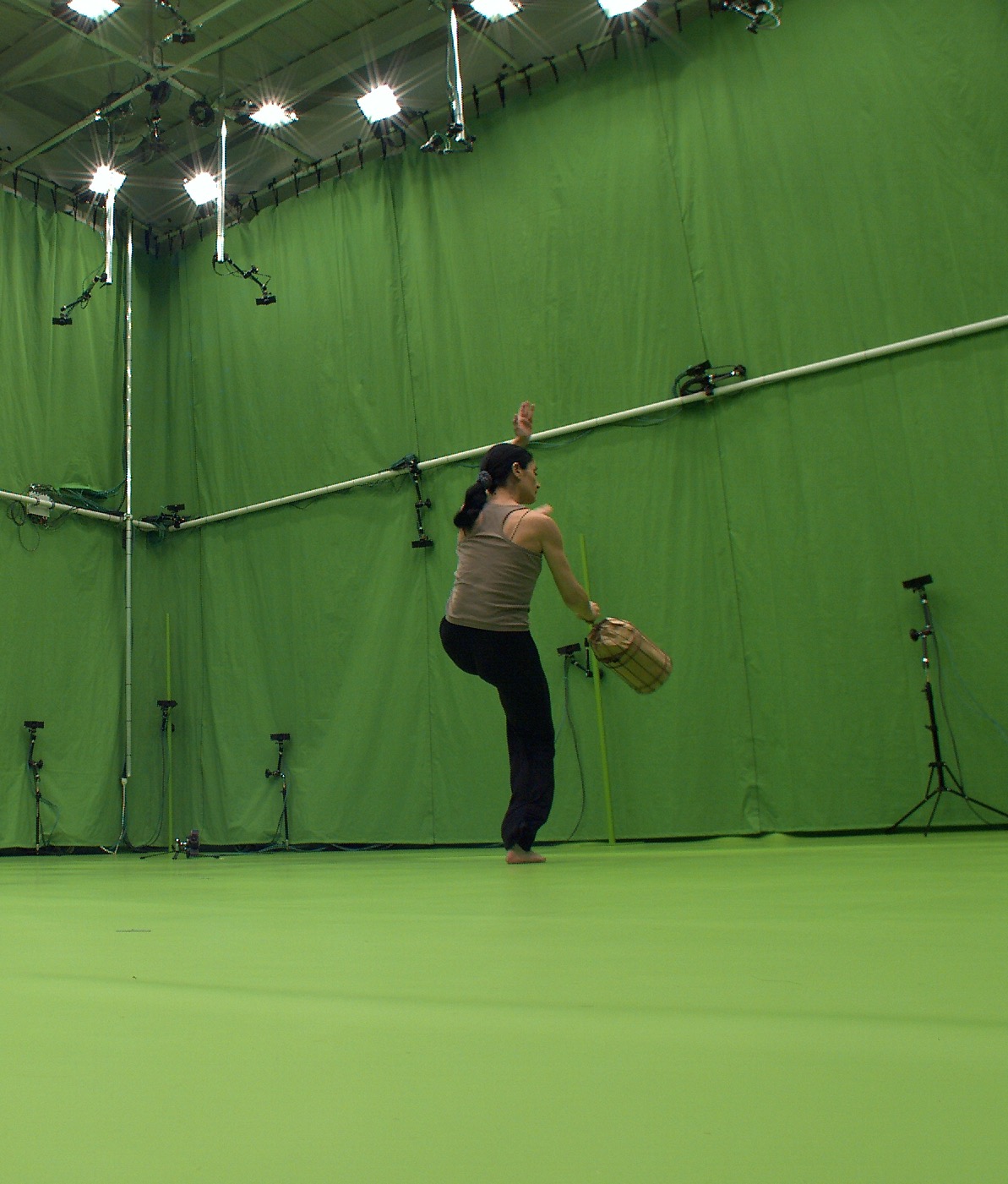} & \includegraphics[width=0.16\linewidth]{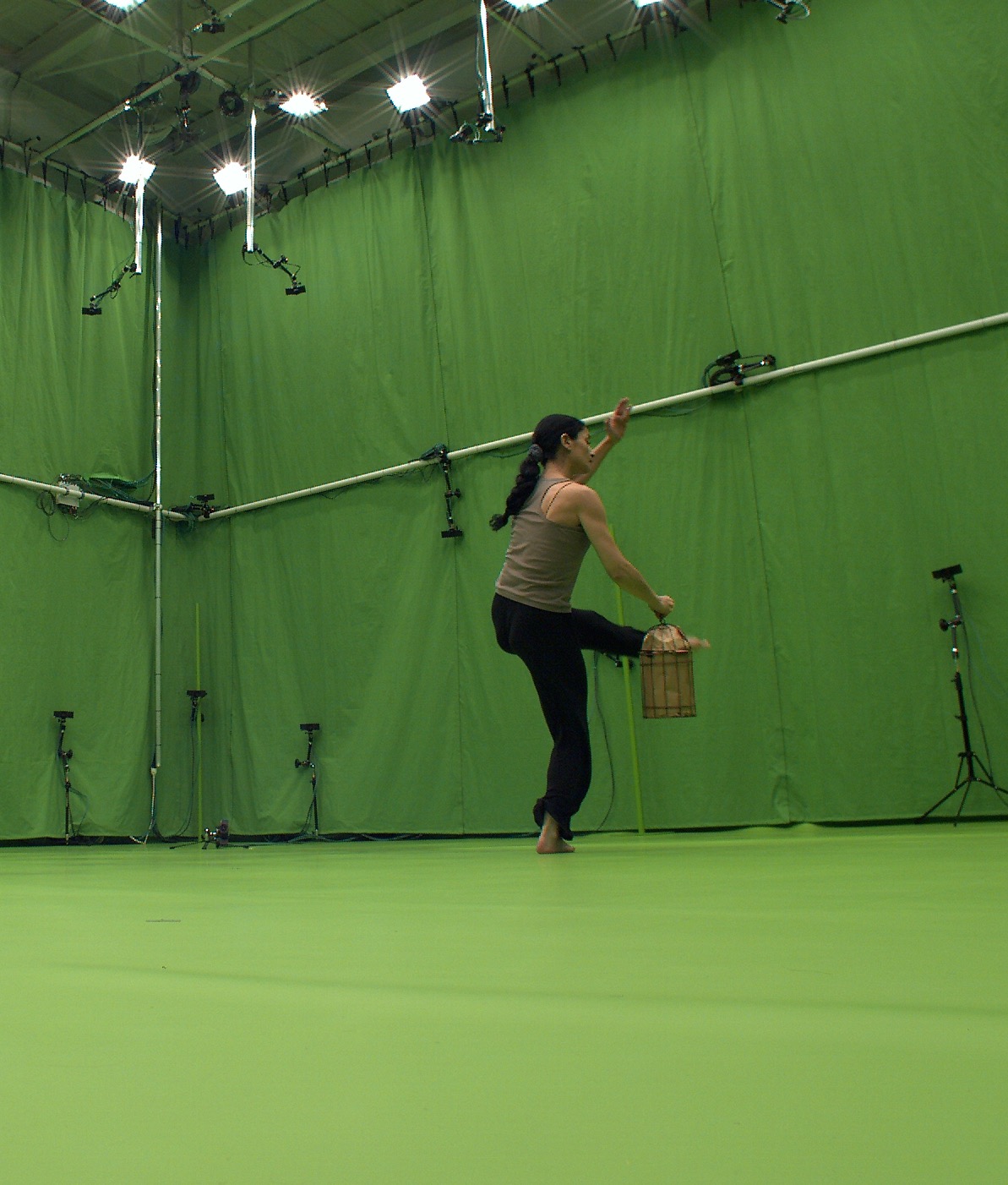}
\end{tabular}
\hspace*{-4mm}
\centering
\begin{tabular} {c}
\includegraphics[width=0.98\linewidth]{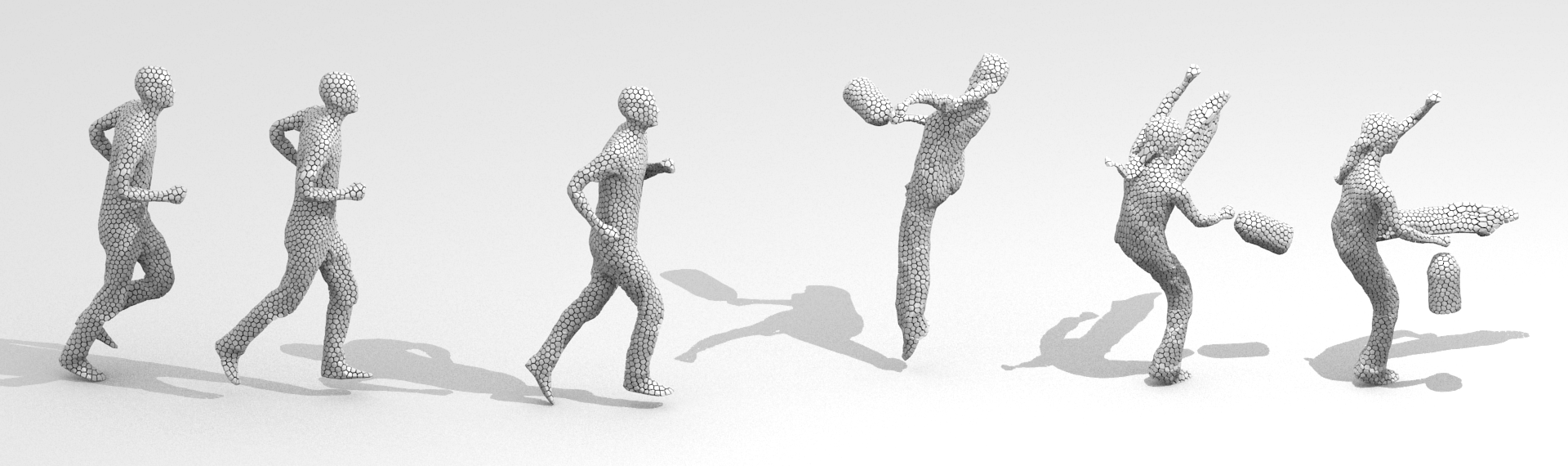}\\
\includegraphics[width=0.98\linewidth]{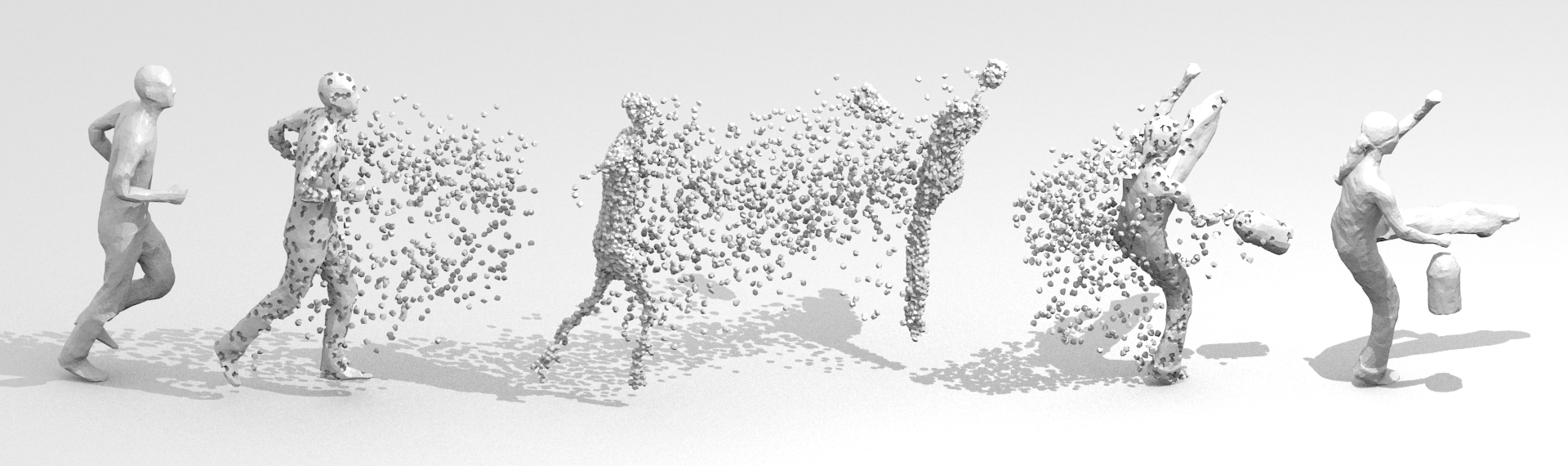}
\end{tabular}
\caption{Tracking result of the \runner{} and the \cagebirddance{} sequences (middle) and combination with volumetric morphing with 5000 cells (bottom).}
\figl{morphing}
\end{figure*}
Our dynamic representations allows to apply volumetric morphing between evolving shapes, enabling therefore new visual effects with real dynamic scenes.  To this purpose, cells of the source shape are first matched to the target shape. Second, each cell is individually morphed to its target cell at a given time within the sequence. Time ordering is chosen such that cells in the source shape are ordered from  the outside to the inside, and associated with the cells of the target shape ordered from the inside to the outside. Cells are transformed from the source to the destination by interpolating their positions and using ~\cite{kent92} to morph their polyhedral shapes.  Figure~\ref{fig:morphing} shows the dynamic morphing of  the \runner{} sequence onto the \cagebirddance{} sequence. 

\begin{figure*}[htbp]
\centering
\hspace*{-4mm}
\begin{tabular} {cc}
\raisebox{.00\height}{\includegraphics[width=0.25\linewidth]{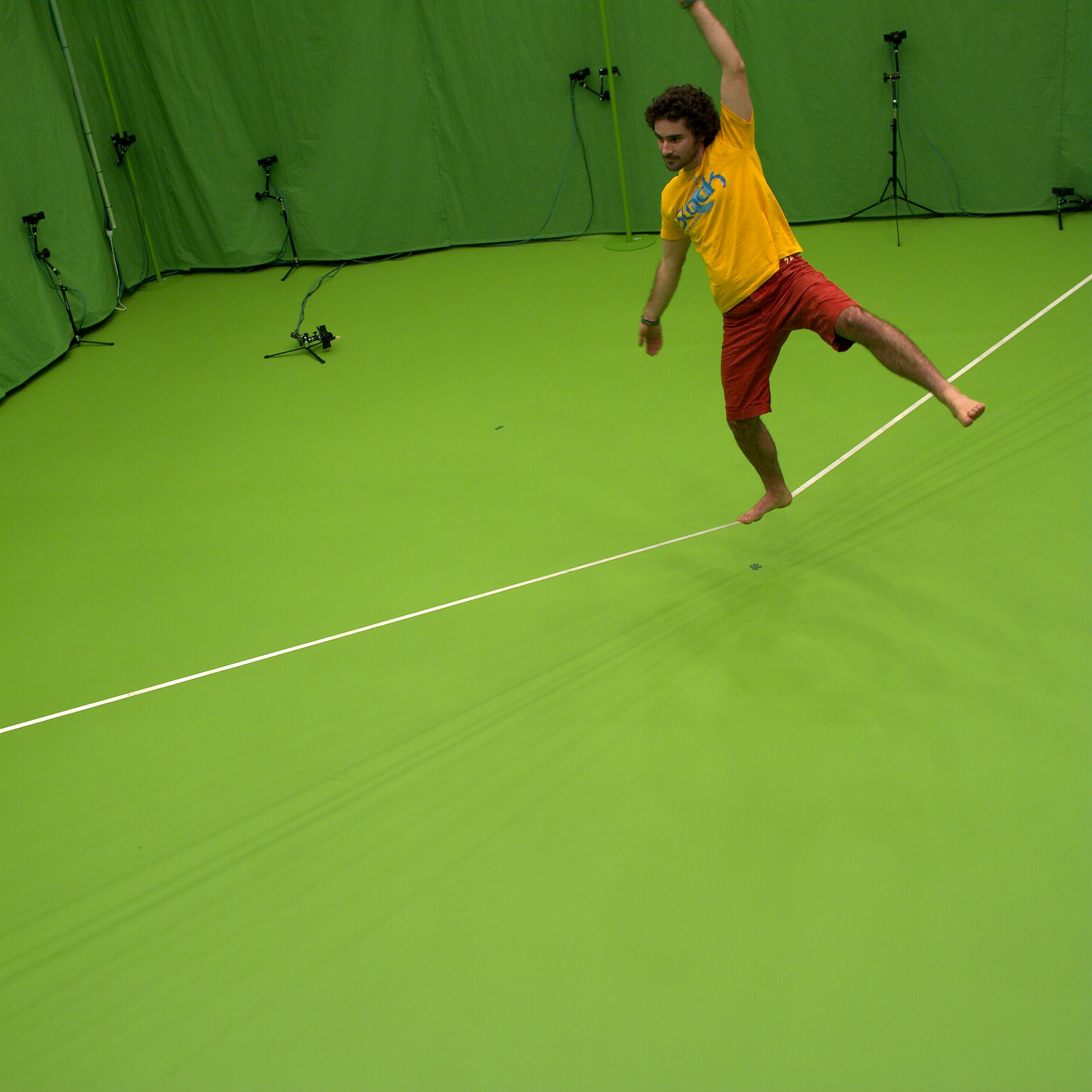} } &
\includegraphics[width=0.7\linewidth]{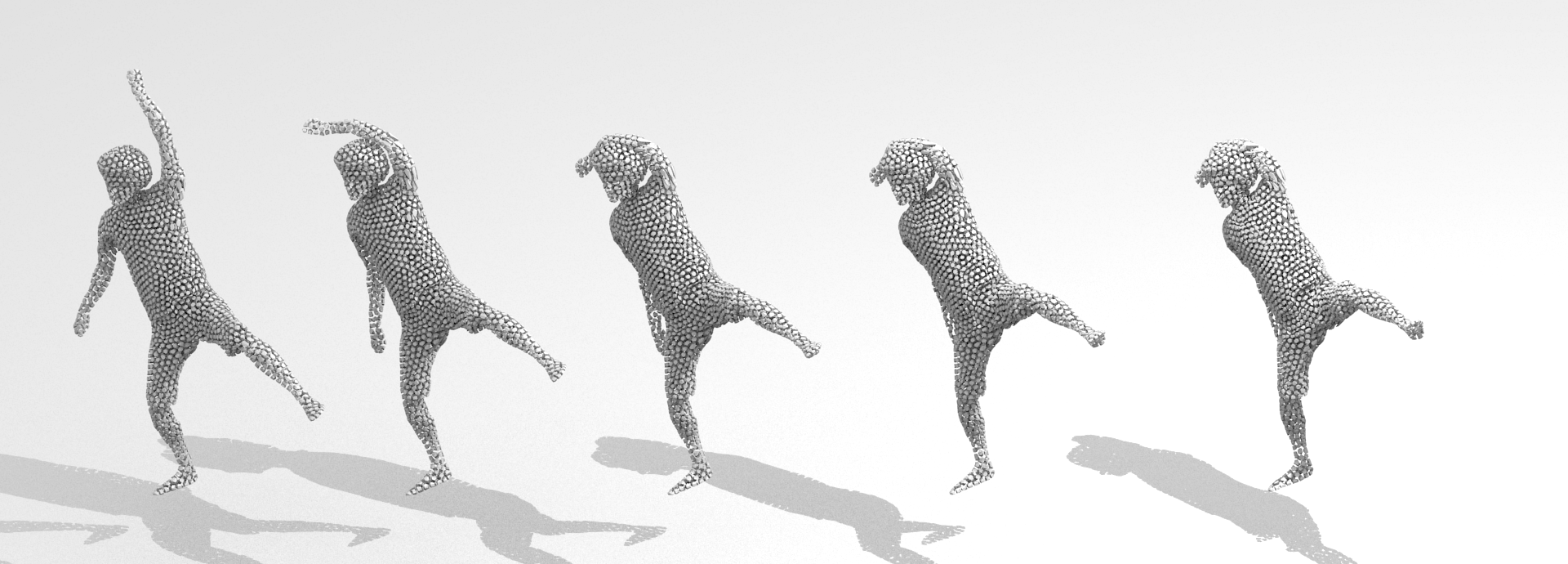}\\
\raisebox{.00\height}{ \includegraphics[width=0.25\linewidth]{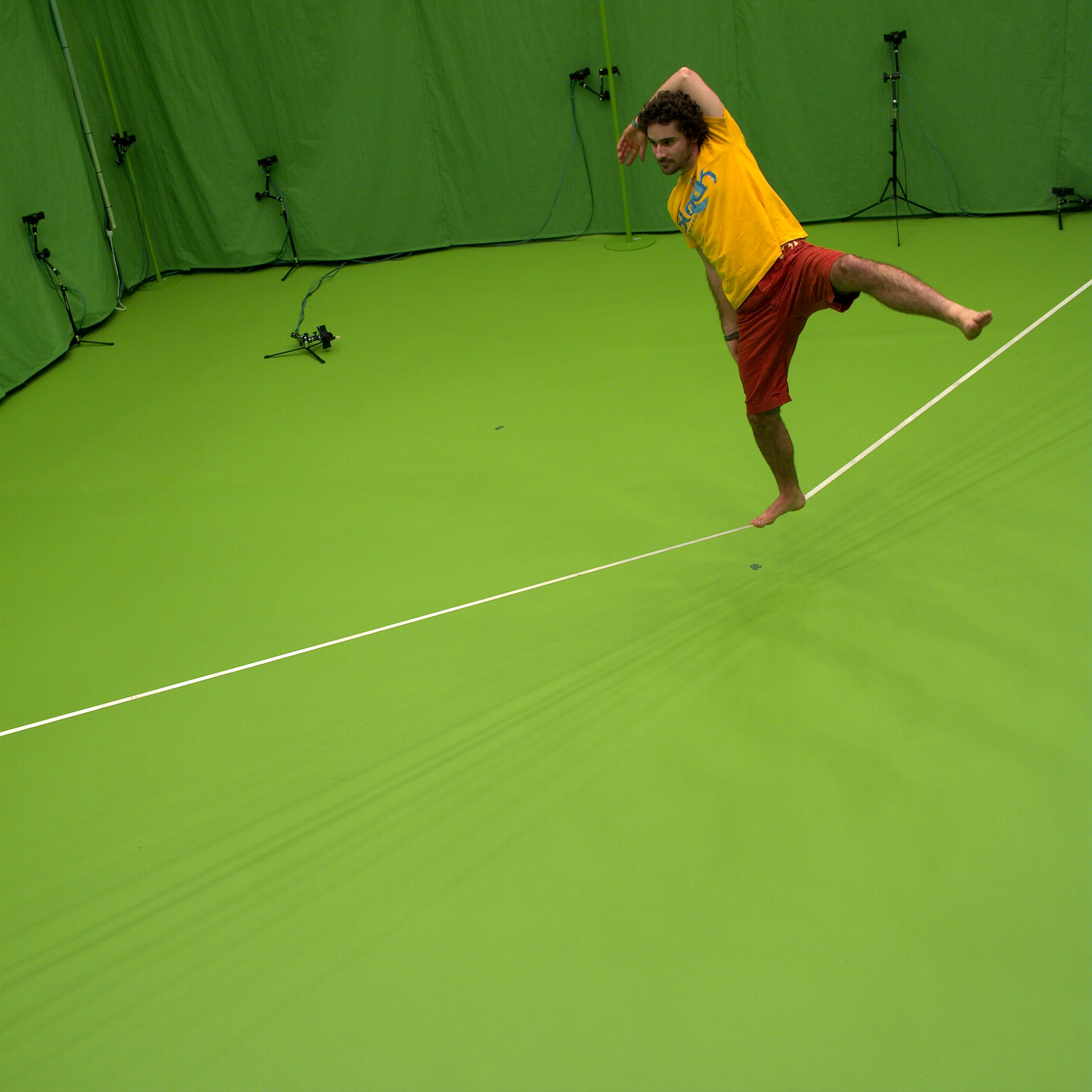} }& 
 \includegraphics[width=0.7\linewidth]{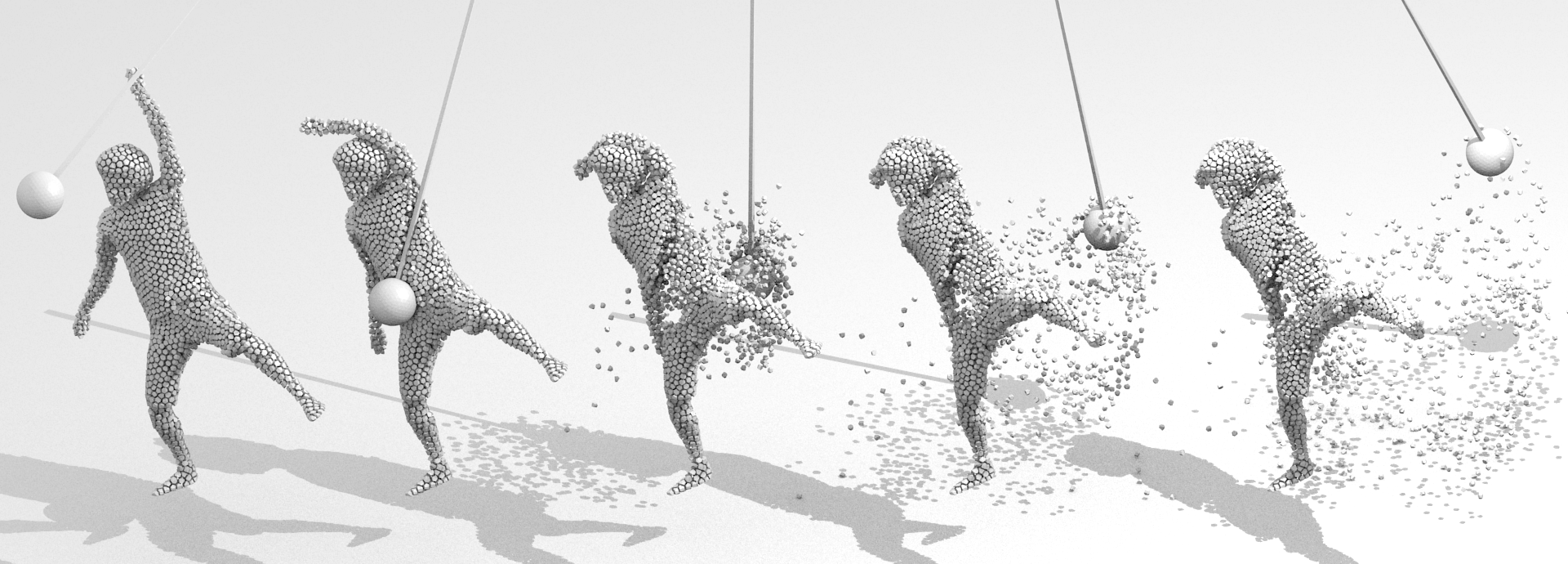}
%\begin{tabular} {c@{\hskip 1mm}c@{\hskip 1mm}c@{\hskip 1mm}c@{\hskip 1mm}c}
%\includegraphics[width=0.19\linewidth]{images/Slackline/img-001007} & \includegraphics[width=0.19\linewidth]{images/Slackline/img-001012} & \includegraphics[width=0.19\linewidth]{images/Slackline/img-001015} & \includegraphics[width=0.19\linewidth]{images/Slackline/img-001017} & \includegraphics[width=0.19\linewidth]{images/Slackline/img-001019}
%\end{tabular}
%\hspace*{-4mm}
%\begin{tabular} {c}
%\includegraphics[width=0.8\linewidth]{images/Slackline/slackline_pendulumA_cut}\\
%\includegraphics[width=0.8\linewidth]{images/Slackline/slackline_pendulumB_cut}
\end{tabular}
\caption{Input \slackline{} Multi-camera observations (left), tracking result of the \slackline{} sequence (top) and combination with the effect of collision with a pendulum (bottom).}
\figl{pendulum}
\end{figure*}

\subsection{Run Time Performance}
\secl{sec:timing}
Our approach has been tested on a dual Intel Xeon E5-2665 processor with 2.40 GHz each.
For each animation, the Poisson runs in 0.67 $s$ per frame on average, and the volumetric decomposition for each frame runs in 6.27 $s$, except for the template model for which it runs in 25.52 $s$ since more iterations are applied.
Our tracking algorithm runs in 45 $s$ per frame on average.
The physical simulation usually needs about 350 $ms$ per simulation step on a single thread. The morphing runs on multiple threads in about 2.35 $s$.

\section{Limitations}
As shown in the previous examples, our approach generates plausible results for a variety of captured and simulated motions. However, a few limitations must be noted.
First, the true captured shape must be volumetric in nature, since we tessellate its interior into 3D cells. Thin shapes such as clothes may cause some cells to be flat, leading to volume variation among cells and ill-defined cohesion constraints that would cause difficulty to the tracking model.

Regarding the physical simulation, our current demonstrator is limited to rigid body interactions, but could be extended to other physical models such as soft body physics and fluid simulation.
Since a CVT provides neighboring information, soft body simulation could be achieved by introducing soft constraints between neighboring cells. This would lead to animations where cell sets behave more like a whole rather than independent bodies.
%Le merge de cellules splittees peut donner autre chose que la forme voulue (?)
%Running time ?

%-------------------------------------------------------------------------
\section{Conclusion}
This document reports on a framework that allows video-based animations to be combined with physical simulation to create  unprecedented and plausible animations. The interest is to take benefit of both modalities and to bring complementary properties to computer animations. Our approach relies on volumetric tessellations within which kinematic constraints can be associated to mechanical forces or procedural tasks at a cell level. Because of its simple and unified nature, relying on centroidal Voronoi tessellations, the system opens various new opportunities to reconsider  how animations can be generated and automated from raw captured 3D data, and it paves the way for other effects such as soft-body and fluid mechanics, which we will explore in future work.

\bibliographystyle{apalike}

\bibliography{augmentedanimation}

\end{document}